\documentclass[10pt, twocolumn, journal]{IEEEtran}
\usepackage{enumerate}
\usepackage{graphicx}
\usepackage{epsf}
\usepackage{subfigure}
\usepackage{amsmath}
\usepackage{amssymb}
\usepackage{array}
\usepackage{setspace}
\usepackage[amsmath,thmmarks,amsthm]{ntheorem}
\usepackage[ruled,lined]{algorithm2e}
\usepackage{algorithmic}
\usepackage{color}
\usepackage{bbm}
\usepackage{cite}
\usepackage{multirow}

\graphicspath{{Fig/},{fig/}}

\theoremseparator{.}

\begin{document}

\title{\Huge Energy Efficient Beamforming Training in Terahertz Communication Systems}

\author{
Li-Hsiang Shen,~\IEEEmembership{Member,~IEEE},
Kai-Ten Feng,~\IEEEmembership{Senior Member,~IEEE},
Lie-Liang Yang,~\IEEEmembership{Fellow,~IEEE}}

\maketitle

\begin{abstract}

	Terahertz (THz) enables promising Tbps-level wireless transmission thanks to its prospect of ultra-huge spectrum utilization and narrow beamforming in the next sixth-generation (6G) communication system. Compared to millimeter wave (mmWave), THz intrinsically possesses compellingly severer molecular absorption and high pathloss serving confined coverage area. These defects should be well conquered under the employment of ultra-thin 3D beamforming with enormous deployed antennas with high beam gains. However, pencil-beams require substantially high overhead of time and power to train its optimal THz beamforming direction. We propose an energy efficient (EE) oriented THz beamforming (EETBF) scheme by separating the original complex problem into beamforming training (EETBF-BT) acquirement and learning-enabled training power assignment (EETBF-PA). The historical beam data is employed to update next beam selection policy. The performance results have demonstrated that the proposed EETBF outperforms the existing benchmarks leveraging full beam search, iterative search, linear/binary search as well as non-power-control based mechanism in open literature. Our proposed EETBF scheme results in the lowest training latency and power consumption, achieving the highest effective rate and EE performance.
	
\end{abstract}


{\let\thefootnote\relax\footnotetext
{Li-Hsiang Shen is with the Department of Communication Engineering, National Central University, Taoyuan 320317, Taiwan. (email: shen@ncu.edu.tw)}}

{\let\thefootnote\relax\footnotetext
{Kai-Ten Feng is with the Department of Electronics and Electrical Engineering, National Yang Ming Chiao Tung University (NYCU), Hsinchu 300093, Taiwan. (email: ktfeng@nycu.edu.tw)}}

{\let\thefootnote\relax\footnotetext
{Lie-Liang Yang is with the Department of Electronics and Computer Science, University of Southampton, Southampton SO17 1BJ, U.K. (e-mail: lly@ecs.soton.ac.uk)
}}

\section{Introduction}

	 With the proliferation of sixth-generation (6G) wireless communications, the tele-traffic demands are explosively escalating due to the emergence of novel applications such as X-reality, ultra-high definition images, and smart big-data \cite{6G1, 6G2, acm,thz_tutorial}. As a paradigm shift from sub-6 GHz and millimeter wave (mmWave), terahertz (THz) \cite{Survey, newnew3} ranging from 0.1 to 10 THz holds the promise of an abundant bandwidth capable of meeting the high data rate requirements of 6G, which attracts considerable attention due to its extremely-huge spectrum utilization. The first THz protocol is ratified by IEEE 802.15.3d \cite{IEEE3d} supporting ultra-wide band up to tens-of-GHz bandwidth at THz, which is applied in the potential stationery scenarios of wireless X-haul, data center, kiosk downloading, and intra-chip networks. Moreover, THz is also employed for mobility use cases \cite{mobile1, mobile2, iot}, such as Internet of nano-mobile things, molecular and nano-sensor/communication mobile networks.

Most of the existing standards and applications focus on the frequency bands below 1 THz having cleaner channels and lower signal losses \cite{RRM2, thz_tutorial}. For example, IEEE 802.15.3d \cite{IEEE3d} standard operates at around 300 GHz, providing data rates up to 100 Gbps over short distances. These frequencies are ideal for applications in wireless backhaul, high-speed short-range communication, and imaging systems for security and medical purposes. The frequency band from 0.3 THz to 0.5 THz is adopted for ultra-high-speed communication for 6G development, enabling applications such as holographic projections and joint communication and sensing. The frequency band from 0.6 THz to 1 THz is also regarded as part of future 6G communications systems. The European Union's Horizon 2020 programs \cite{euro}, such as TeraNova and WORTECS, are exploring these frequencies for Tbps transmissions. However, the higher frequency bands from 2 THz to 3 THz, are not yet widely utilized for communication purposes. Instead, these frequencies are primarily applied in fields of spectroscopy and astronomy, where they are particularly suitable for scientific research, e.g., molecular physics and deep-space observations.

	 While, THz physical properties impose several implementation hurdles with respect to both the high power signal attenuation and detection, which requires advanced crafted THz beamforming techniques \cite{Survey}. Due to much smaller wavelengths compared to mmWave, THz is capable of utilizing ultra massive multi-input-multi-output (UM-MIMO) \cite{UMMIMO1,UMMIMO2,newnew2} to generate narrow three-dimensional (3D) beams offering considerably-high antenna gains under miniaturized manufactured on-device space. Different from conventional mmWave networks, THz arises potential issues in physical and medium access control layers, including high-frequency channel modeling \cite{Measure}, molecular absorption modeling \cite{Measure2}, pencil-beam orientation \cite{CH3}, and power consumption \cite{UMMIMO2}. The current research papers primarily concentrate on evaluating the feasibility of THz beam-based channel measurement as shown in \cite{CH1,CH2,CH3}, combating the issues of short transmission distances. As an enhancement from conventional hybrid beamforming, the paper \cite{UMMIMO2, newnew1} has rolled out an advanced beamforming from hardware architecture perspective, which employs array-of-subarrays with each carrying out different functionalities for spectrum- and energy-efficient (EE) solutions. Under such architecture, beam control becomes compellingly imperative. The authors in \cite{RRM1} have designed an appropriate long-term THz beam control for dynamic wireless power transfer in mobile nano-cells. In paper \cite{RRM2}, an EE nano-node beam association is designed by considering the energy limitation and THz channel properties, whilst quality-of-service is additionally constrained in \cite{RRM2-1}. With the THz narrow-beam control, the paper \cite{RRM3} elaborates the benefits of spatial diversity of THz beamforming in non-orthogonal multiple access. In \cite{RRM4}, the authors have designed distance-aware adaptive THz beamforming, which manages system power and antenna subarray selection, i.e., beam selection. Other studies are exploring the usage of metasurfaces to enhance the coverage area of THz communications \cite{RIS1,RIS2,RIS3}. However, this induces additional expenditure in terms of further complex optimization of surface configuration and deployment, higher power consumption, and expenses related to manufacturing as well as maintenance. Moreover, the above-mentioned works of \cite{UMMIMO2,RRM1, RRM2, RRM2-1, RRM3, RRM4, RIS1, RIS2, RIS3} are designed based on perfect beamforming, which neglect the overhead and defect of beamforming training before desirable data transmission \cite{IEEE3d}, which potentially deteriorates the performance of pencil-beam-oriented THz systems.

	 In a generic protocol ratified by IEEE 802.11ad/ay \cite{Myad1, IEEEaday} for mmWave as well as in IEEE 802.15.3d \cite{IEEE3d} for THz, a frame, without loss of generality, is partitioned into two sessions, including beamforming training and data transmission durations. Excessive training overhead leaves little time for desirable data transfer, whilst overlength framing gives rise to outdated beam/channel information, especially in a short-distance THz network with mobility. In papers \cite{RIS1, RIS1-1}, the authors have designed a hierarchical tree-based THz beam training or so-called iterative search, which is a more advanced mechanism compared to that in mmWave \cite{mmWTree}. However, the global ground truth of narrow-beam index cannot be readily obtained owing to different beam patterns in a tree-based training. This has been well-addressed in the works of \cite{Myad1,THzTree1, THzTree2} by given a threshold guaranteeing adequate beam signal strength. Moreover, the initial stage of beamforming tree involves searching through wide beamwidths, which may inadvertently detect nearby nodes because of comparably low beam-gain as well as received power than narrow-beams. This situation has not been addressed or resolved in the papers of \cite{RIS1, RIS1-1, THzTree1, THzTree2}. In \cite{IEEE3d}, exhaustive beam training is performed, which requires considerable overhead and potentially leaves no time for data transmission in a UM-MIMO enabled THz system. The authors in \cite{THzBT} have designed a low-overhead beam training method using detailed physical layer information. Similar concept in mmWave can be transferred to THz domain by employing only beam index and averaged signal strength at higher layers for massive beam training \cite{Myad2, Myad3}. To elaborate a little further, linear/binary beam search \cite{mmWLBS} is conventionally utilized but requiring frequent information feedback, whilst non-feedback based mobility-aware beamforming is conceived by \cite{Myad4}. However, most of the existing works in \cite{mmWTree, RIS1, RIS1-1, THzTree1, THzTree2, IEEE3d, THzBT, mmWLBS} do not consider the compelling tradeoff among metrics of beam signal quality, beam alignment accuracy and training latency as well as throughput, which should be jointly designed as pointed out by \cite{Myad1}. In this context, high-efficiency THz beam training and alignment have been regarded as the highly challenging tasks, which should be considered when designing THz beamforming.

As previously explained, during beamforming training phase, weaker beams with low beam-gain can be compensated by extending the frame time, which may lead to a lack of time for data transfer. Moreover, all configured beams may become infeasible impinged by environmental pitfall, regardless of the length of training frame. To mitigate this problem, we may utilize higher power during beamforming training to compensate for implausible beams. On the other hand, the energy can be preserved under sufficiently-strong beams, achieving high-EE training. Nevertheless, to the best of  our knowledge, none of the existing works \cite{mmWTree, RIS1, RIS1-1, THzTree1, THzTree2, IEEE3d, IEEEaday, THzBT, mmWLBS, Myad1, Myad2, Myad3, Myad4} have taken the dynamic power control into account during beamforming training. Instead, they always consume full-power, which potentially leads to energy wasting, weaker beams, and even unreachable signals. It is worth noting that the training power control is quite different from the instantaneous power allocation used for maximizing data transmission, as discussed in \cite{RRM1, RRM2, RRM3, RRM4}. During training, the power should be determined before acquiring information on either channels or beams, which is unknown and stochastic. This process substantially relies on probabilistic optimization as well as historical beamforming data \cite{Myad1}. Furthermore, with the rapid development of powerful machine and deep learning  techniques, reinforcement learning has found to be in favour of solving complex problems in the dynamically-changing environments in the fields of wireless communications and networking \cite{RL2, RL3, add1, add2}. In reinforcement learning \cite{RL}, an agent interacts with the environment and updates the model based on the corresponding rewards, which is potentially capable of achieving a long-term optimal policy. Therefore, by utilizing the historical beam results and incorporating the learning-assisted power control, we design a low-complexity EE-oriented THz beamforming approach, which substantially strikes a compelling tradeoff between latency and accuracy as well as effective throughput/EE. The contributions of this paper are summarized as follows.
\begin{itemize}
	\item We have conceived joint power control and 3D beamforming training in the EE-oriented THz communication system. It is capable of adjusting the appropriate beams and power for improving EE performance, while satisfying the guaranteed signal quality, available power, allowable training duration and beam alignment accuracy.
	
	\item We propose an EE-oriented THz beamforming (EETBF) scheme, where the original complex problem is partitioned into two sub-problems, namely acquiring candidate 3D beamforming training (EETBF-BT) and enabling dynamic training power assignment (EETBF-PA) through reinforcement learning. The proposed EETBF tends to extend the training duration in the cases of low beam alignment, whilst it maintains the training power to ensure acceptable signal quality. Historical beam data is employed to update the subsequent beam policy. Moreover, initial beam selection is adopted based on the channel quality obtained from historical beams. 	
	
	\item We guarantee a polynomial time complexity with the aid of long-term learning optimization. The performance results have demonstrated that the proposed EETBF outperforms the existing benchmarks leveraging full beam search, iterative search, linear/binary search as well as the non-power-control based mechanism found in open literature. It achieves the lowest training latency and power consumption as well as the highest effective rate and EE performance.	

\end{itemize}
	
	The rest of this paper is organized as follows. In Section \ref{SM}, we present the THz system model, which includes bemforming, channel modeling, and beam training, followed by the problem formulation. The proposed EETBF approach adopting 3D beamforming training and learning-enabled training power control is elaborated in Section \ref{BP}. Performance evaluations of proposed EETBF scheme are conducted in Section \ref{PE}. Finally, conclusions are drawn in Section \ref{CON}.
	
\begin{figure}[!t]
\centering
	\includegraphics[width=3.3 in]{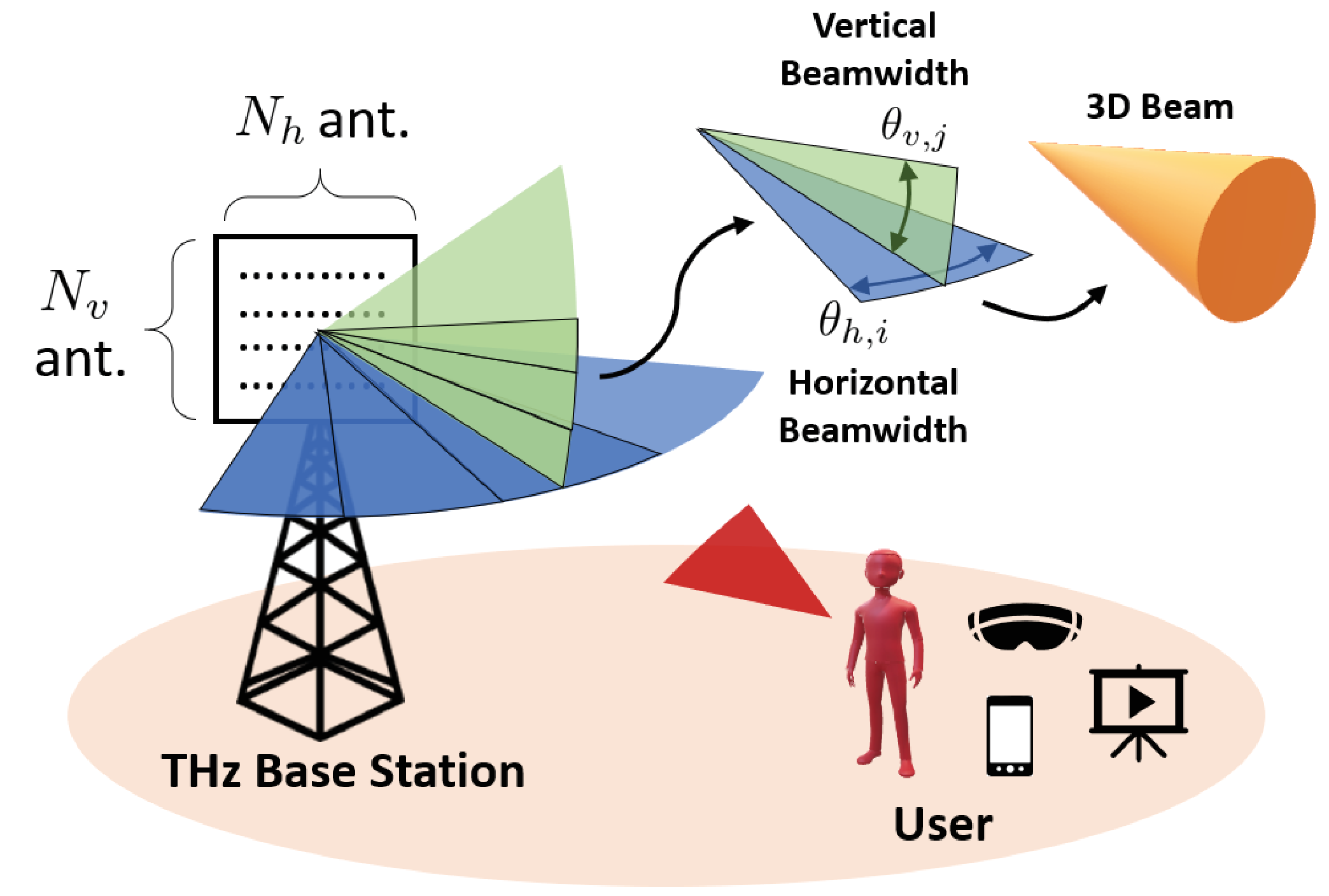}
	\caption{Architecture of THz BS deployed with an antenna array having $N_{h}$ horizontal and $N_v$ vertical antenna elements. The corresponding beamwidths of the $i$-th horizontal and the $j$-th vertical beams are $\theta_{h,i}$ and $\theta_{v,j}$, respectively.}
	\label{scene}
\end{figure}
	
\section{System Model and Problem Formulation} \label{SM}

\subsection{Beamforming Model}
	As shown in Fig. \ref{scene}, we consider a THz base station (BS) deployed with an antenna array with a total number of $N_{t}=N_{h}\cdot N_v$ antennas, where $N_{h}$ and $N_v$ denote the numbers of horizontal and vertical antenna elements, respectively. The THz BS located at $\boldsymbol{X}_{ap}(0,0,h)$ is capable of conducting THz 3D beamforming to serve the desired moving user positioned at $\boldsymbol{X}_{ue}(x,y,z)$ within its serving coverage with the corresponding distance $d(t)=\left[ x^2+y^2 + \left(h-z\right)^{2}\right]^{1/2}$ between the THz BS and user. The beam profile is defined as the beam direction $\psi_{\mathcal{A},x}$ and beamwidth $\theta_{\mathcal{A},x}$, where $\mathcal{A}\!\in\!\{h,v\}$ denotes the horizontal and vertical beamforming directions, respectively, and the subscript $x\!=\!\{i,j\}$ indicates the index of either the $i$-th horizontal or the $j$-th vertical beam. It is noted that the THz beam is generated using full antennas to compromise the intrinsically severe path loss of THz transmissions. The beamforming gain of either horizontal or vertical beam direction is proportional to the configured number of antennas, where the THz beamforming gain can be asymptotically acquired as 
\begin{align} \label{bg}
 G_{\mathcal{A}}\left(\theta_{\mathcal{A},x}, \psi_{\mathcal{A},x} \right) \approx 
\left\{\begin{array}{ll}
	\frac{2\pi}{\theta_{\mathcal{A},x}}, \, \mbox{ if } |\psi_{\mathcal{A},x} - \tilde{\psi}_{\mathcal{A}}| \leq \frac{\theta_{\mathcal{A},x}}{2},\\
	\varepsilon,  \quad\quad  \mbox{otherwise}.
\end{array} \right.
\end{align}
From \eqref{bg}, we can infer that the THz mainlobe beam in the first expression of $\eqref{bg}$ is that the directional beam angle $\psi_{\mathcal{A},x}$ falls into the boresight angles $\tilde{\psi}_{\mathcal{A}}$ of the THz BS, which are acquired as $\tilde{\psi}_{h} = \tan^{-1} (y/x)$ and $\tilde{\psi}_{v} = \tan^{-1} (d/(z-h))$, respectively. The corresponding beamwidth is further computed as $\theta_{\mathcal{A},x} = 1.772\pi d_{t}/N_{\mathcal{A}}$, where $d_{t}$ is the broadening factor determined by antenna spacing. Notation $\varepsilon \ll 1$ is a generalized constant gain for sidelobes. Note that the proposed system model can be theoretically extended to the bi-directional beams in both transmitter and receiver ends. However, considering practical cases, the user may only be equipped with a few antennas due to implementation issues. Therefore, we will neglect the receiving beamforming in the following discussions.
	
\subsection{THz Channel}
	Inspired by \cite{RRM2, RRM4, Measure2}, the distance-aware path loss of THz channel fading can be expressed as
\begin{align} \label{ch}
	H_{i,j}(t,f,d) = \rho_{i,j}(t)\left( \frac{c}{4 \pi f d} \right)^{2} e^{-\mathbbm{K}_{T\!H\!z}(f) d},
\end{align}
where $\rho_{i,j}(t)$ is the small-scale variation of 3D beam $(i,j)$ at time instant $t$, $f$ is THz operating frequency, $c$ is a constant of speed light, and $\mathbbm{K}_{THz}(f)$ is the molecular absorption coefficient related to frequency. Notice that both the effects of line-of-sight (LOS) and non-light-of-sight (NLOS) THz channel can be asymptotically modeled by $\eqref{ch}$ \cite{RRM2}. As stated in \cite{Measure2}, different THz channels will induce distinct coefficients of $\mathbbm{K}_{T\!H\!z}(f)$ based on the experimental measurement, which requires table mapping. We can readily observe from $\eqref{ch}$ that THz experiences additional absorption loss as decaying exponentially, which provokes compellingly high channel attenuation when compared to microwave and mmWave communications. Moreover, the environmental THz noise led by molecular perturbation is expressed as \cite{Measure2}
\begin{align} \label{noise}
	N(f,d) = k_{B} T_{0} \left(1-e^{-\mathbbm{K}_{T\!H\!z}(f) d}\right),
\end{align}
where $k_{B}$ is the Boltzmann constant and $T_{0}$ is the reference temperature. It is worth mentioning that the noise model of \eqref{noise} differs from that in micro/millimeter waves, since THz is substantially influenced by molecules alternating its native physical properties under different transmission distances \cite{RRM4}.

\begin{figure*}[!t]
\centering
\subfigure[]{\includegraphics[width=1.5 in]{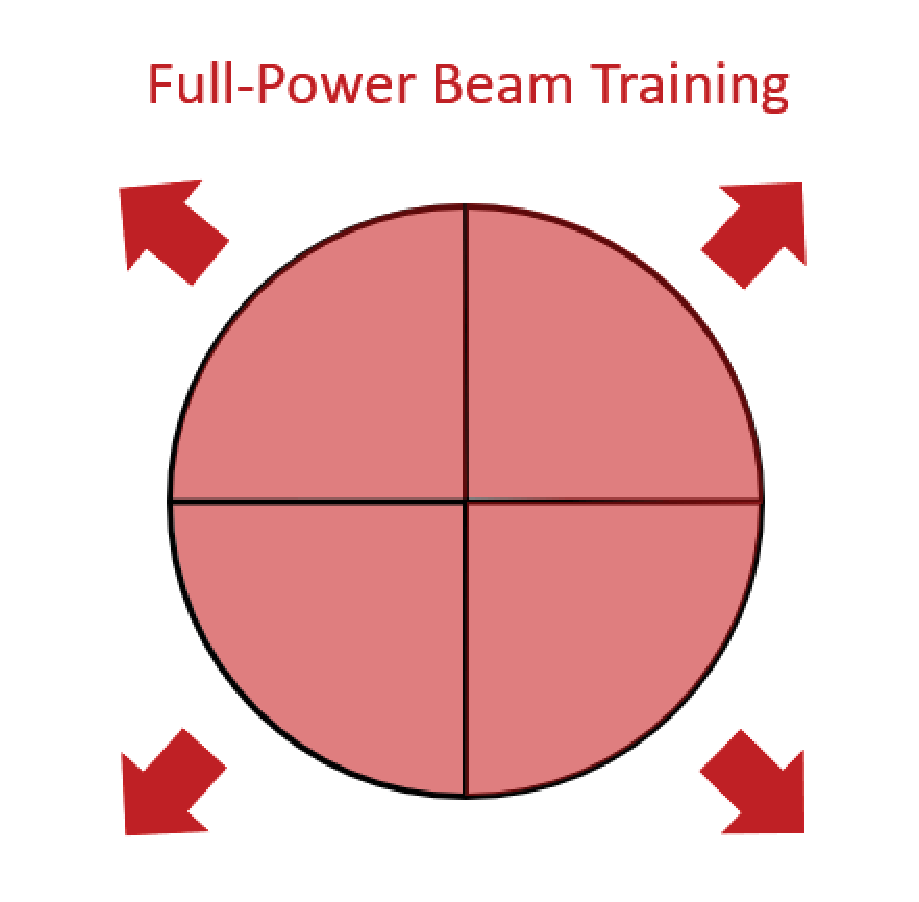} \label{bt1}}
\subfigure[]{\includegraphics[width=1.7 in]{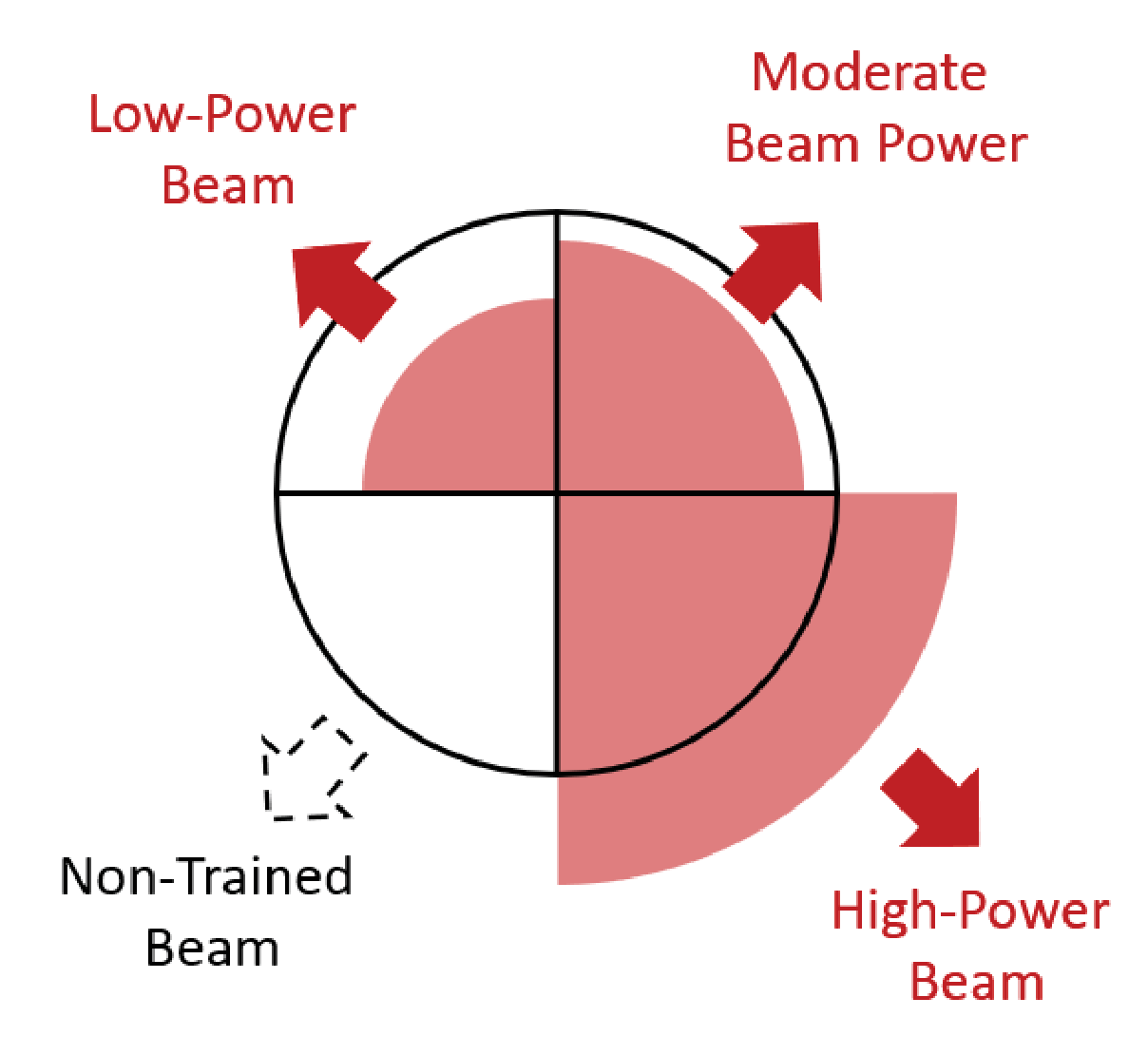} \label{bt2}}
\subfigure[]{\includegraphics[width=3.3in]{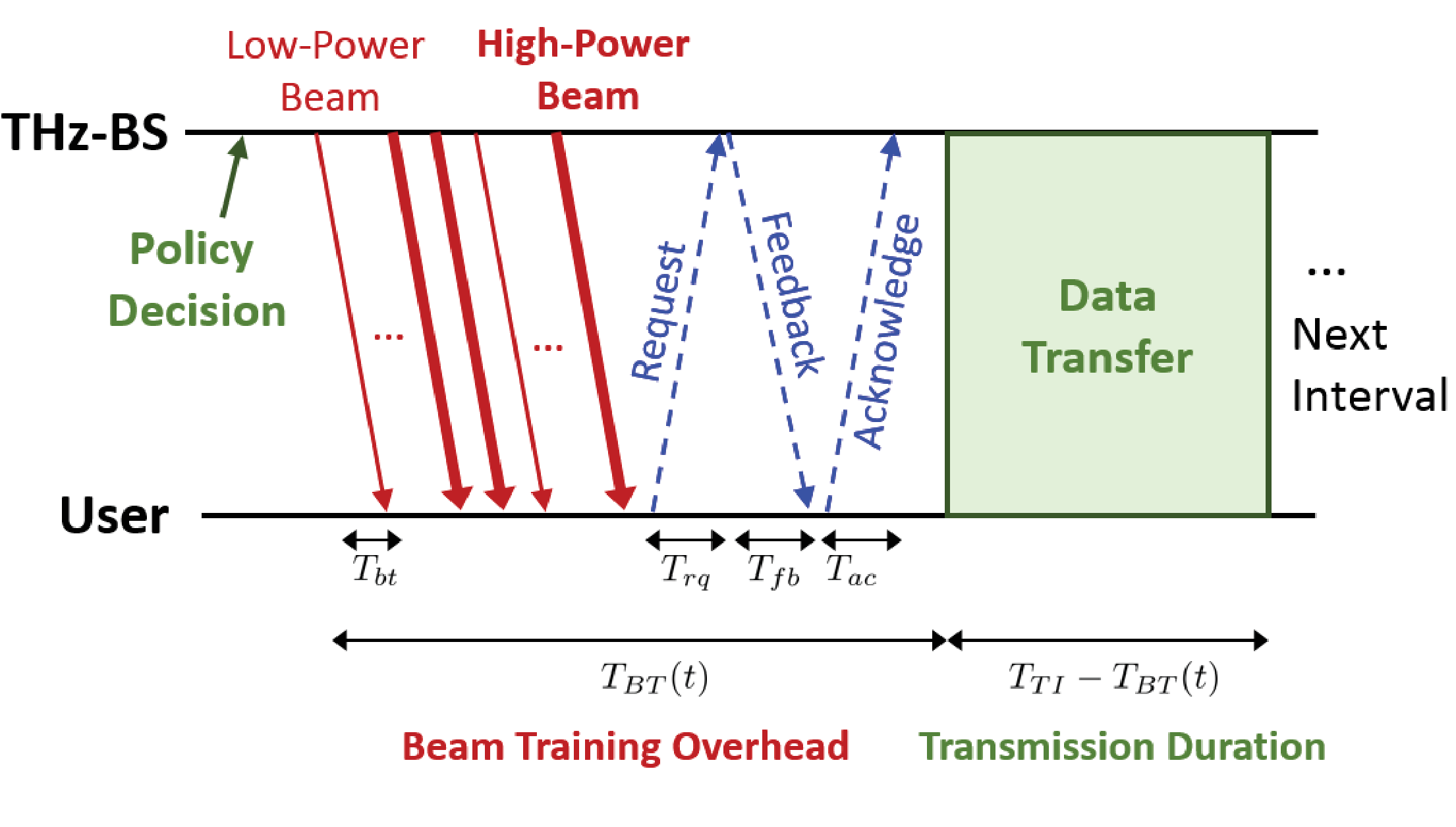}\label{protocol}}
\caption{The beam training model and framework. (a) Conventional full-power beam training. (b) Proposed EE-oriented beam training. (c) The designed protocol of THz beamforming training, including beam training overhead and transmission duration. Note that the policy decision is performed before initialiation of beam training.}
\label{protocol_all}
\end{figure*}


\subsection{THz Beamforming Training}
	In Fig. \ref{protocol_all}, we illustrate the beam training model and the corresponding framework. We can observe that the conventional beamforming training in Fig. \ref{bt1} performs full-beam search as well as full-power training, which might lead to wasted energy and abundant training overhead. By contrast, in our proposed model shown in Fig. \ref{bt2}, we consider EE-oriented beam training, which relies on high power to compensate for low-quality beams and low power to preserve energy for strong beams. Moreover, 
partial beamforming training is conducted since some beams may become less informative. The overall procedure of beamforming training is demonstrated in Fig. \ref{protocol}. We consider that the THz-BS is configured with the available predefined codebook for THz 3D beam sector set of $\boldsymbol{\Phi} = \{\boldsymbol{\Phi}_{h}, \boldsymbol{\Phi}_{v}\}$ with the candidate 3D beam training set of $\boldsymbol{\phi}(t) = \{\boldsymbol{\phi}_{h}(t),\boldsymbol{\phi}_{v}(t)\}$ at the $t$-th interval. The BS is assumed to train a single 3D beam using a single time slot with the signal-to-noise ratio (SNR) of the $(i,j)$-th beam received by the user at frame $t$, which can be represented by
\begin{align} \label{sinr}
\gamma_{i,j}(t) = \frac{P_{i,j}(t) H_{i,j}(t,f,d) \cdot G_{h}(\theta_{h,i},\psi_{h,i})\cdot G_{v}(\theta_{v,j},\psi_{v,j})}{N(f,d) + N_{N\!F}},
\end{align}
where $\mathbf{P}(t)=\{P_{i,j}(t)| \forall i\!\in\!\boldsymbol{\phi}_{h}(t), \forall j\!\in\! \boldsymbol{\phi}_{v}(t)\}$ is the training power set, and $N_{N\!F}$ indicates the power of additional noise figure from hardware devices. Based on the fundamental THz beamforming training process in \cite{IEEE3d} that is inherited from the similar concept in \cite{Myad1}, the user will provide feedback on the received SNR $\gamma_{i,j}(t)$ to the THz BS with the optimal beam index as
\begin{align}\label{ind}
	k^{*}(t) = \left\lbrace i^{*}(t), j^{*}(t)\right\rbrace = \underset{{i\in\boldsymbol{\phi}_{h}(t), j\in\boldsymbol{\phi}_{v}(t)}}{\arg\!\max} \gamma_{i,j}(t).
\end{align}
The BS will then store the optimal beam index $k^{*}(t) = \left\lbrace i^{*}(t), j^{*}(t)\right\rbrace$ with the corresponding SNR value of $\gamma_{i^*,j^*}(t)$. Owing to practical implementation of storage constraint, we assume that the THz-BS can store historical beam dataset from the past $M$-interval optimal beam indexes and corresponding SNR values, which is denoted as $\boldsymbol{\Xi}(t,M)=\{ k^{*}(t-\tau), \gamma_{i^*,j^*}(t-\tau)|\forall 1\leq\tau\leq M\}$.

	As for training overhead, we define $T_{bt}$ as the duration of beam training for a single THz beam, which is followed by the request and feedback as well as acknowledgement phases with the respective durations of $T_{rq}$, $T_{fb}$ and $T_{ac}$. We can therefore acquire the total overhead in terms of beamforming training latency at interval $t$, given by
\begin{align}\label{lat}
T_{BT}(t)= |\boldsymbol{\phi}(t)|\cdot T_{bt}+T_{rq}+T_{fb}+T_{ac}.
\end{align}

Consider that arbitrary beamforming training should be explicitly completed within a transmission interval $T_{TI}$, i.e., $T_{BT}(t)\leq T_{TI}$; otherwise, there will be no session for data transfer, leading to zero throughput. Based on the aforementioned definitions, the achievable effective throughput when considering training overhead is expressed as
\begin{align}\label{rt}
R(t) &= W \cdot  \max\Big(0,1-\alpha(t)\Big) \cdot \notag \\
& \qquad\qquad \log_{2}\Big( 1+ \gamma_{i^*,j^*}(t) \mathbbm{1}\{\gamma_{i^*,j^*}(t)\geq \gamma_{th} \} \Big),
\end{align}
where $W$ is the system bandwidth, and $\alpha(t)=T_{BT}(t)/T_{TI}$ is the ratio of THz beam training overhead. The $\max(\cdot, \cdot)$ prevents negative throughput, meaning that beam training duration exceeds the given transmission interval. In $\eqref{rt}$, the received SNR of the optimal beam is guaranteed to be higher than the predefined threshold of $\gamma_{th}$, which is regarded as the successful training, where $\mathbbm{1}\{\cdot\}$ is an indicator function. 

The averaged power consumption comprises both beamforming training and data transmission power which can be represented by
\begin{align} \label{power}
&P(t) = \notag \\
& \frac{1}{T_{TI}}\underbrace{\left( \sum_{i\in \boldsymbol{\phi}_{h}\!(t)} \sum_{j\in \boldsymbol{\phi}_{v}\!(t)} T_{bt}P_{i,j}(t) \right)}_{\text{Beam Training Power}} 
	+ \underbrace{\left(T_{TI}-T_{BT}\right) P_{i^{*},j^{*}}(t)}_{\text{Data Transmission Power}}.
\end{align}
Therefore, the effective EE leveraging THz beamforming training latency in $\eqref{lat}$, rate performance in $\eqref{rt}$ and the power consumption in $\eqref{power}$ can be acquired as
\begin{align} \label{ee}
EE(t)=\frac{R(t)}{P(t)}.
\end{align}
Since THz communications are substantially affected by high path loss compared to microwave and mmWave, it becomes compellingly imperative to dynamically adjust the beamforming training power to achieve high EE, which however has not been considered in existing literature. Moreover, as the globally optimal beam may be unattainable, we accordingly characterize the averaged beam training alignment accuracy with acceptably-high SNR as
\begin{align} \label{acc}
\xi(t) = \frac{1}{M}\sum_{\tau=0}^{M} \mathbbm{1}\left\lbrace\gamma_{i^*,j^*}(t-\tau) \geq \gamma_{th}\right\rbrace.
\end{align}

\subsection{Problem Formulation}
Based on the above analysis, in this paper, our design aims at acquiring an appropriate policy of the beam training candidate set and corresponding training power under the maximization of the system effective EE and the consideration of joint beamforming training power, throughput, latency overhead, and beam alignment accuracy as follows:
\begingroup
\allowdisplaybreaks
\begin{subequations} \label{OPT}
  \begin{align} 
    &\underset{\mathbf{P}(t),\boldsymbol{\phi}(t)}{\text{max}} \quad EE(t)
    \label{Obj}\\
    &\text{s.t.} \quad \gamma_{i^*,j^*}(t)\geq \gamma_{th}, \quad \forall i^*\in\boldsymbol{\phi}_{h}(t), j^*\in\boldsymbol{\phi}_{v}(t), t,  \label{C1}\\
    &\quad\quad \gamma_{i,j}(t)\geq \gamma_{dec}, \quad\ \forall i\in\boldsymbol{\phi}_{h}(t), j\in\boldsymbol{\phi}_{v}(t), t,  \label{C1-1}\\
    &\quad\quad P_{i,j}(t) \leq P_{be}, \, \quad\ \forall i\in\boldsymbol{\phi}_{h}(t), j\in\boldsymbol{\phi}_{v}(t), t, \label{C2}\\
    &\quad\quad P(t) \leq P_{th},\quad\quad\ \forall t, \label{C3}\\
    &\quad\quad T_{BI}(t)\leq T_{TI},\quad\ \forall t, \label{C4}\\
    &\quad\quad \xi(t)\geq \xi_{th}\qquad\quad\ \forall t. \label{C5}
  \end{align}
\end{subequations}
\endgroup
In the above problem, $\eqref{Obj}$ jointly considers the THz 3D beamforming training regarding the latency, throughput and power tradeoff. In this regards, we note that this is the first work jointly studying these issues. The beam signal is constrained by the trained optimal beam SNR in $\eqref{C1}$, whereas all SNR values of training beams should be decodably higher than $\gamma_{dec}$ in $\eqref{C1-1}$. The allowable power $P_{be}$ per beam direction is given in $\eqref{C2}$ and the available total power of THz-BS $P_{th}$ is provided in $\eqref{C3}$. We also confine the allowable training time length in $\eqref{C4}$ as well as the required beam alignment accuracy threshold $\xi_{th}$ in $\eqref{C5}$. We can infer from $\eqref{OPT}$ that it consists of the mixed discrete integer and continuous variables, which give a nonlinear and non-convex problem. Moreover, the solution to the problem is intrinsically undetermined and stochastic, since the optimal beam performance is potentially unattainable until the completion of the full beamforming training procedure. Accordingly, we conceive to partition the problem into the beam selection and power assignment sub-problems by proposing the implementable schemes to approximately solve the EE maximization problem $\eqref{OPT}$.

\section{Proposed Energy Efficient THz Beamforming (EETBF)} \label{BP}

\subsection{Proposed EETBF Scheme}
	Due to the non-convexity and nonlinearity in the EE problem $\eqref{OPT}$, we partition it into two sub-problems, namely the beamforming training $\boldsymbol{\phi}(t)$ of EETBF-BT and the power adjustment $\mathbf{P}(t)$ of EETBF-PA, where the corresponding BT and PA sub-problems are formulated as
  \begin{align} \label{OPT1}
    &\underset{\boldsymbol{\phi}(t)}{\text{max}} \quad EE(t) \quad \text{s.t.} \ \eqref{C1}, \eqref{C1-1}, \eqref{C4}, \eqref{C5},
  \end{align}
and
  \begin{align} \label{OPT2}
    &\underset{\mathbf{P}(t)}{\text{max}} \quad EE(t)  \quad \text{s.t.}\ \eqref{C1} \text{--} \eqref{C5}.
  \end{align}
According to the protocol designed in Fig. \ref{protocol}, the beam strength is unattainable until the completion of the whole training process. Therefore, the only information we have is the stored historical beam dataset of $\boldsymbol{\Xi}(t,M)$, which is employed to obtain the optimal policy followed by power adjustment. The overall algorithm is elaborated as Algorithm \ref{Alg1}. We require initialization of beam configuration $\boldsymbol{\Phi}$ and $\boldsymbol{\phi}(t)$, time counter $t$ as well as performance indicators, such as latency, power consumption, throughput, beam alignment accuracy and EE. The exhaustive search with maximum power is conducted to initialize the dataset collection for $M$ periods, whereas EETBF-BT/PA schemes are executed from the $(M+1)$-th interval. The details of both sub-schemes are elaborated in the following  subsections.

\begin{algorithm}[!tb] \label{Alg1}
\caption{\small Proposed EETBF Scheme}
\SetAlgoLined
\DontPrintSemicolon
\small
\begin{algorithmic}[1]
\STATE Initialization: \\	
	1) Configure THz BS 3D beam with horizontal/vertical beam set $\boldsymbol{\Phi}=\{\boldsymbol{\Phi}_{h}, \boldsymbol{\Phi}_{v}\}$ and training beam set $\boldsymbol{\phi}(t)=\{\boldsymbol{\phi}_{h}(t),\boldsymbol{\phi}_{v}(t)\}$\\
	2) Initialize training latency $T_{BT}(t)$, power consumption $P(t)$, throughput $R(t)$, alignment accuracy $\xi(t)$, and EE $E(t)$\\
	3) Set time counter $t=0$\\

\REPEAT
	\STATE Transmission interval increment as $t=t+1$
	\IF {$t<M$ or insufficient accuracy $\xi(t)<\xi_{th}$}
		\STATE Exhaustive beam training $\boldsymbol{\phi}(t)=\boldsymbol{\Phi}$\\
		\STATE Configure the maximum power $P_{i,j}(t)=P_{be}$
	\ELSE
		\STATE Perform Algorithm \ref{Alg2} to obtain horizontal and vertical beam training set $\boldsymbol{\phi}(t)=\{\boldsymbol{\phi}_{h}(t),\boldsymbol{\phi}_{v}(t)\}$ 
		\STATE Conduct Algorithm \ref{Alg3} to acquire power allocation result $\mathbf{P}(t)=\{P_{i,j}(t)|\forall i\in\boldsymbol{\phi}_{h}(t), \forall j\in \boldsymbol{\phi}_{v}(t)\}$
	\ENDIF
	\STATE Obtain beamforming set $\boldsymbol{\phi}(t)$ by spanning horizontal and vertical beam index sets $\{\boldsymbol{\phi}_{h}(t),\boldsymbol{\phi}_{v}(t)\}$ 
	\STATE THz BS executes 3D beamforming training employing beam policy $\boldsymbol{\phi}(t)$ and power assignment $P_{i,j}(t)$
	\STATE User feedbacks the optimal beam indexed by $k^{*}(t) = \left\lbrace i^{*}(t), j^{*}(t)\right\rbrace$ in $\eqref{ind}$ with SNR $\gamma_{i^*, j^*}(t)$\\
	\STATE The THz BS stores the beam result in $\boldsymbol{\Xi}(t,M)$
	\STATE Compute performances of training latency $T_{BI}(t)$, effective throughput $R(t)$, and EE $EE(t)$ given by $\eqref{lat}$, $\eqref{rt}$ and $\eqref{ee}$, respectively.
	\STATE Update Q-table in Algorithm \ref{Alg3}
\UNTIL{Termination of beamforming}

\end{algorithmic}
\end{algorithm}

\begin{algorithm}[!tb] \label{Alg2}
\caption{\small EETBF-BT}
\SetAlgoLined
\DontPrintSemicolon
\small	
\begin{algorithmic}[1]
\STATE Initialization: \\
	1) Configure horizontal/vertical beam $\boldsymbol{\Phi}_{\mathcal{A}}$, $\boldsymbol{\phi}_{\mathcal{A}}(t)$ where $\mathcal{A}=\{h,v\}$ \\
	2) Given historical observation beam dataset $\boldsymbol{\Xi}(t,M)$ with window size $M$, SNR threshold $\gamma_{th}$\\

		\STATE Obtain historical optimal beam indexes $\boldsymbol{\phi}'(t)=\{x^*(t-\tau) | \forall x=\{i,j\}, \tau=1,2,...,M\}$\\
		\STATE Compute difference of beam index difference $D_{\mathcal{A}}={\rm diff}_{\mathcal{A}}(\boldsymbol{\phi}'(t))$\\
		\STATE Set spanning size $s_{\mathcal{A}}=1$\\
		\FOR{$\tau=1,...,M$}
			\IF{$\gamma_{i^{*}, j^*}(t-\tau)\leq\gamma_{th}$}
				\STATE Increase size $s_{\mathcal{A}}=s_{\mathcal{A}}+1$ due to failure training in history\\
			\ELSE
				\STATE Learn central beam index $\psi_{\mathcal{A}} = \boldsymbol{\phi}'(t-s) $ and spanning width $\varepsilon_{\mathcal{A}} = s_{\mathcal{A}}\cdot\max\,  |D_{\mathcal{A}}|$ and terminate counting\\
			\ENDIF
		\ENDFOR		
		\IF {$\min \, D_{\mathcal{A}}\geq0$}
			\STATE Forward search as $\boldsymbol{\phi}_{\mathcal{A}}(t)=\{\psi_{\mathcal{A}},\psi_{\mathcal{A}}+1,...,\psi_{\mathcal{A}}+\varepsilon_{\mathcal{A}}\}$\\
		\ELSIF {$\max \, D_{\mathcal{A}}<0$}
			\STATE Backward search as $\boldsymbol{\phi}_{\mathcal{A}}(t)=\{\psi_{\mathcal{A}}-\varepsilon_{\mathcal{A}}, \psi_{\mathcal{A}}-\varepsilon_{\mathcal{A}}+1,...,\psi_{\mathcal{A}}\}$\\
		\ELSE
			\STATE Bi-directional search as $\boldsymbol{\phi}_{\mathcal{A}}(t)=\{\psi_{\mathcal{A}}-\varepsilon_{\mathcal{A}},\psi_{\mathcal{A}}-\varepsilon_{\mathcal{A}}+1,...,\psi_{\mathcal{A}}+\varepsilon_{\mathcal{A}}\}$\\
		\ENDIF
		\STATE \textbf{Output}: Beam policy $\boldsymbol{\phi}_{\mathcal{A}}(t)$

\end{algorithmic}
\end{algorithm}

\begin{algorithm}[!tb] \label{Alg3}
\caption{\small EETBF-PA}
\SetAlgoLined
\DontPrintSemicolon
\small	
\begin{algorithmic}[1]
\STATE Initialization:\\
\STATE {\bf (Training Power Acquirement)}\\
	\STATE Quantize power policy into $Q$ levels
	\STATE Randomize a small number $\delta=\left[0,1\right]$ \\
	\IF{$\delta\leq\delta_{th}$}
		\STATE Randomize a power action $a_t$ from discrete power set $\mathcal{P}$\\
	\ELSE
		\STATE Select optimal power action from Q-table as $a_t=\arg\!\max_{a'} Q(s_t,a')$\\
	\ENDIF
	\STATE Mapping power from action by $P_{i,j}(t)=a_t\cdot P_{be}/Q$ and compute total power $P(t)$ by $\eqref{power}$
	\IF{$P(t) > P_{th}$} 
		\STATE Normalize individual power as $P_{i,j}(t)=P_{i,j}(t)\cdot P_{th}/P(t)$\\
	\ENDIF
	\STATE \textbf{Output}: Training power assignment $P_{i,j}(t)$

\STATE {\bf (Q-table Update)}\\
\STATE Update current state by $s_{t+1}=\mathcal{R}\left(\frac{R(t)\cdot s_Q}{R_{max}} \right)$\\
\STATE Update Q-table by $Q(s_t,a_t)\leftarrow Q(s_t,a_t) + \eta_1\left[ EE(t) + \eta_2 \cdot \max_{a'}(s_{t+1}, a') - Q(s_t, a_t) \right]$\\

\end{algorithmic}
\end{algorithm}

\subsubsection{EETBF-BT}

\begin{figure*}[!t]
\centering
\subfigure[]{\includegraphics[width=2 in]{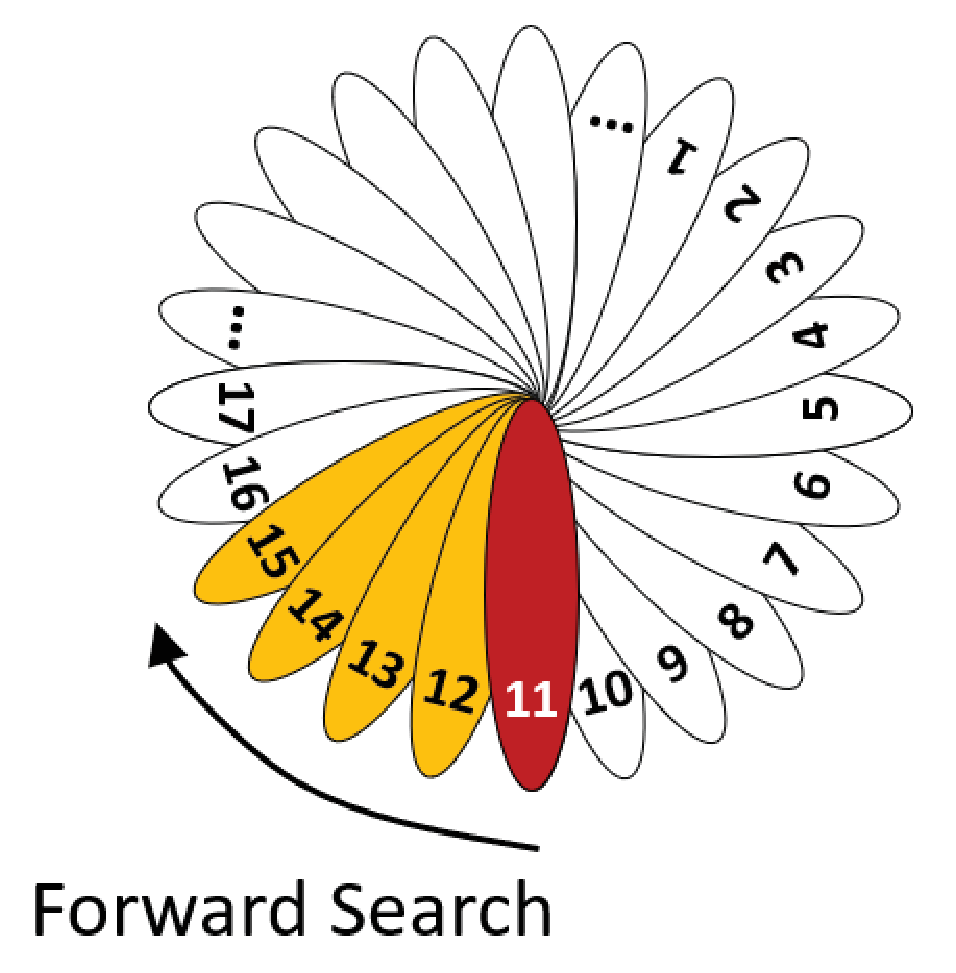} \label{par1}}
\subfigure[]{\includegraphics[width=2 in]{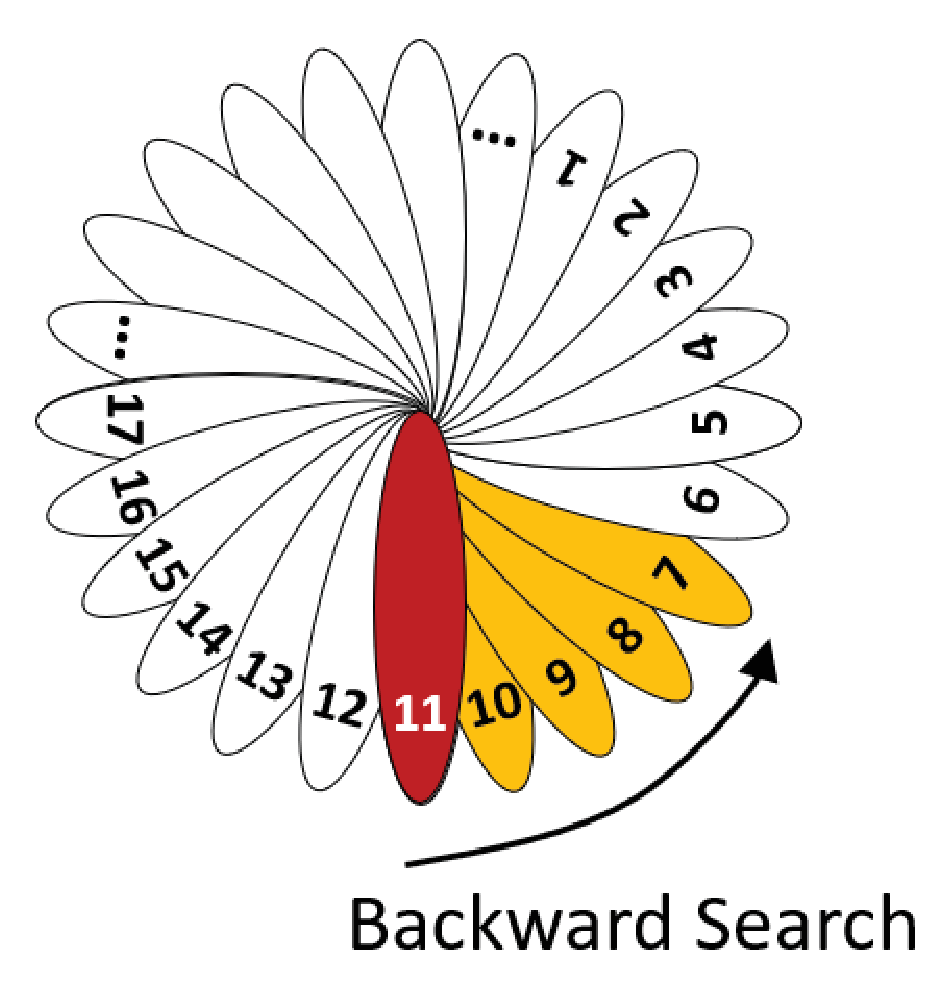} \label{par2}}
\subfigure[]{\includegraphics[width=2 in]{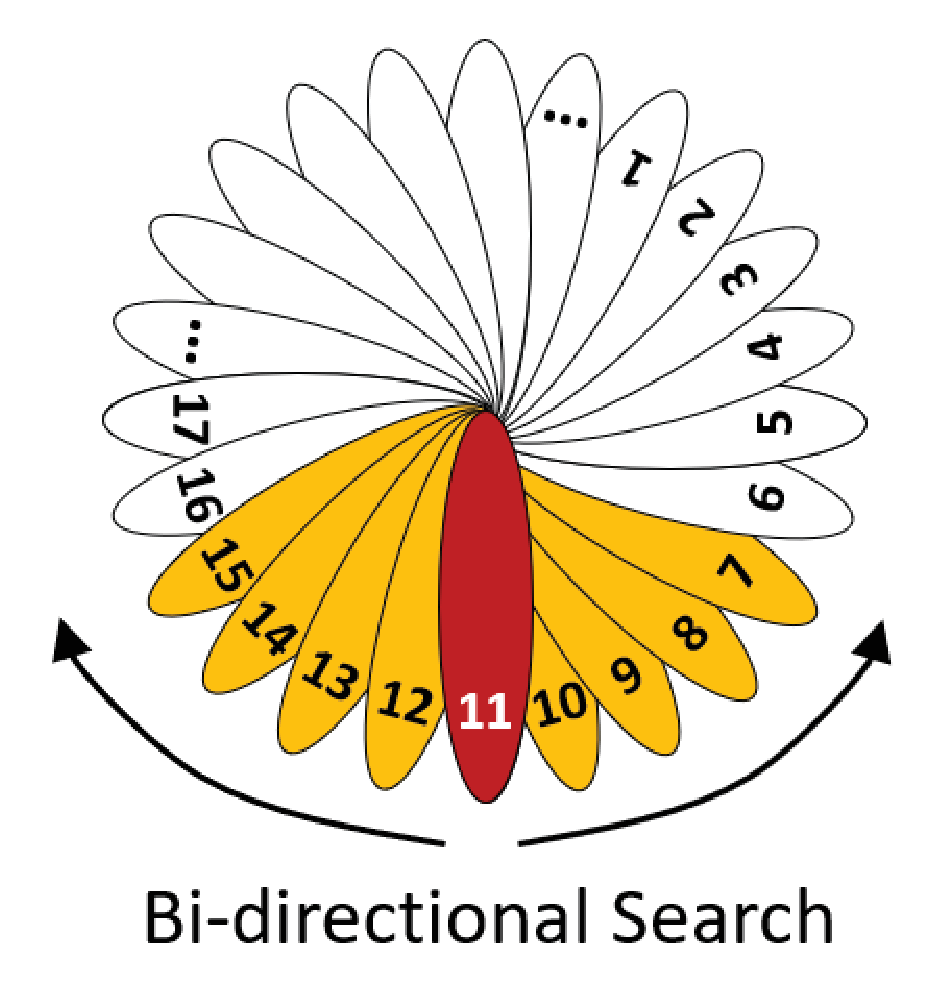} \label{par3}}
\subfigure[]{\includegraphics[width=6 in]{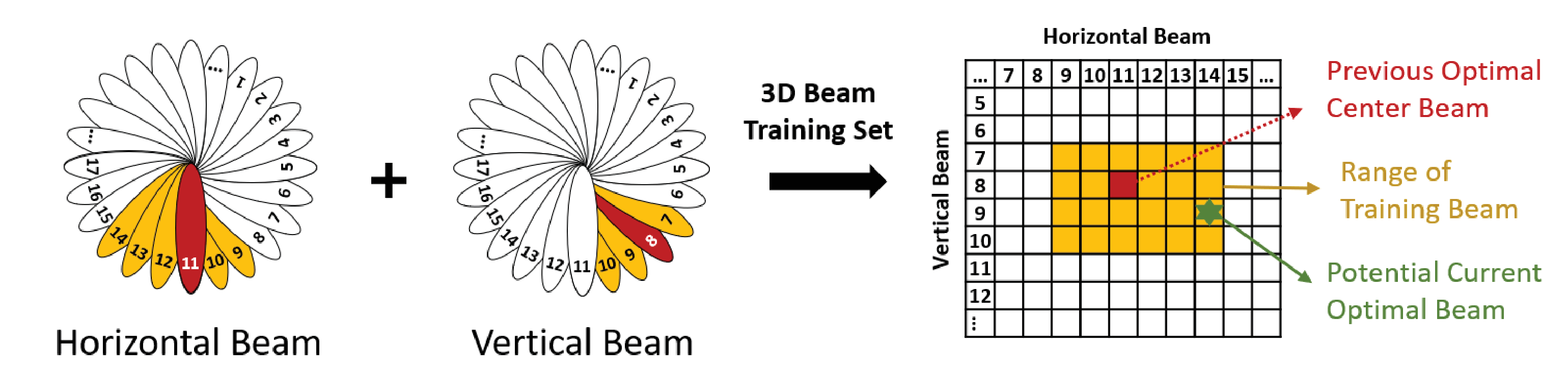} \label{par4}}
\caption{Illustration of the beamforming training regarding (a) forward search in $\eqref{s1}$, (b) backward search in $\eqref{s2}$ and (c) bi-directional search in $\eqref{s3}$. (d) The final 3D beamforming training set. Note that the beam with dark red indicates the previous optimal beam index $\psi_{\mathcal{A}}$, whereas the beams with bright yellow represent the expanded search area calculated based on $\eqref{sss}$.}
\label{partial_beam}
\end{figure*}

Since 3D THz beamforming is considered, we should determine both the horizontal and vertical beam policies as $\{\boldsymbol{\phi}_{h}(t),\boldsymbol{\phi}_{v}(t)\}$. Inspired by \cite{Myad4}, previous results may possess certain correlation with the current unknown and stochastic decision. Moreover, as shown in Algorithm \ref{Alg2}, we independently process both dimensional beams and combine them finally by multiplying the two dimensions together. Given the historical beam dataset $\boldsymbol{\Xi}(t,M)$ with window size $M$ and SNR threshold $\gamma_{th}$, we can then derive the differences of beam index by $D_{\mathcal{A}}={\rm diff}_{\mathcal{A}}(\boldsymbol{\phi}'(t))$, where historical optimal beam indexes are defined as $\boldsymbol{\phi}'(t)=\{x^*(t-\tau) | \forall x=\{i,j\}, \tau=1,2,...,M\}$. We denote the search spanning size $s_{\mathcal{A}}$ indicating the number of timeslots of encompassing consecutive failures in previous beamforming training. If a worse beam with low SNR is acquired, i.e., $\gamma_{i^{*}, j^*}(t-\tau)\leq\gamma_{th}$ is regarded as the failure training in history, we iteratively enlarge the search size by $s_{\mathcal{A}} = s_{\mathcal{A}}+1$. By contrast, we terminate to learn the central beam index as $\psi_{\mathcal{A}} = \boldsymbol{\phi}'(t-s_{\mathcal{A}})$ with a total spanning width $\varepsilon_{\mathcal{A}} = s_{\mathcal{A}}\cdot\max\,  |D_{\mathcal{A}}|$. In the followings, we determine searching direction in three cases:
\begin{subequations} \label{sss}
	\begin{align}
		\boldsymbol{\phi}_{\mathcal{A}}(t)&=\{\psi_{\mathcal{A}},\psi_{\mathcal{A}}+1,...,\psi_{\mathcal{A}}+\varepsilon_{\mathcal{A}}\},  \notag \\
		&\qquad\rightarrow \text{forward search if } \min \, D_{\mathcal{A}}\geq0 \label{s1}, \\
		\boldsymbol{\phi}_{\mathcal{A}}(t)&=\{\psi_{\mathcal{A}}-\varepsilon_{\mathcal{A}}, \psi_{\mathcal{A}}-\varepsilon_{\mathcal{A}}+1,...,\psi_{\mathcal{A}}\}, \notag \\
		&\qquad\rightarrow \text{backward search if } \max \, D_{\mathcal{A}}<0 \label{s2}, \\
		\boldsymbol{\phi}_{\mathcal{A}}(t)&=\{\psi_{\mathcal{A}}-\varepsilon_{\mathcal{A}},\psi_{\mathcal{A}}-\varepsilon_{\mathcal{A}}+1,...,\psi_{\mathcal{A}}+\varepsilon_{\mathcal{A}}\}, \notag \\
		&\qquad\rightarrow \text{otherwise as bi-directional search} \label{s3}.
	\end{align}
\end{subequations}
Note that "forward/backward" indicates relative directions with the increment/decrement of horizontal and vertical beam indexes. In Fig. \ref{partial_beam}, we provide an example illustrating the beamforming training in $\eqref{sss}$. Note that here we only show the beam indexes in either horizontal or vertical dimension. We consider that the previous optimal central beam with its index given by $\psi_{\mathcal{A}}=11$ and its corresponding search widths of $\varepsilon_{\mathcal{A}}=4$. In Fig. \ref{par1}, if $\min \, D_{\mathcal{A}}\geq0$, we perform forward search and its beam training index set is given by $\{11,12,13,14,15\}$. If $\max \, D_{\mathcal{A}}<0$ takes place, we execute backward search with the beam index set of $\{7,8,9,10,11\}$ in Fig. \ref{par2}. In Fig. \ref{par3}, the worst case of bi-directional search occurs when the above conditions are not satisfied, resulting in a beam search set of $\{7,8,9...,13,14,15\}$. After the completion of the beam index set determination in both dimensions, we proceed to combine both horizontal and vertical beam training sets. For example, as depicted in Fig. \ref{par4}, we have determined the horizontal beam set $\{9,10,...,13,14\}$ and the vertical one $\{7,8,9,10\}$. Accordingly, the 3D beam training set is generated through combining the indexes from both dimensions.

\subsubsection{EETBF-PA}
For conventional optimization methods, the power assignment is conducted under known estimated channel information. Nonetheless, in the beam training case, no information is attainable except the historical outcomes. Consequently, the policy becomes undetermined and stochastic. Moreover, different beam directions potentially represent specific environmental conditions leading to difficulties in power adjustment. In this context, we employ reinforcement learning, as a long-term optimal policy, to intelligently and dynamically allocate the beamforming training power, which is elaborated in Algorithm \ref{Alg3}. Note that the deep learning based reinforcement learning is deemed to be inappropriate for such implementation due to its high overhead of model training and unattainable input state of channel information. The important concept of reinforcement learning is the interaction between an agent and the environment, where an action $a_t$ will be selected according to the resulting reward $r_t$ and current state $s_t$ in the established Q-table $Q(a_t, s_t)$. The reward is defined as a function of effective EE $r_t=EE(t)$. Furthermore, we define a quantized power set $\mathcal{P}=\{i\cdot P_{be}/Q | \forall 1\leq i\leq Q\}$ with action $a_t=i$ denoted as an index of a single power solution, where we map the final power decision from action via 
\begin{align}\label{Qp}
	P_{i,j}(t)=a_t \cdot P_{be}/Q.
\end{align}
Notice that $\eqref{Qp}$ implicitly fulfills the constraint in $\eqref{C2}$ since the maximum power value occurs when $a_t=i=Q$, i.e., $P_{i,j}(t)=P_{be}$. If the total power is beyond the predefined budget in $\eqref{C3}$, we conduct normalization utilizing $P_{i,j}(t)= P_{i,j}(t)\cdot P_{th}/P(t)$. Note that the state at the next interval $t+1$ is updated by leveraging the current rate result $R(t)$ as
\begin{align}\label{Qs}
	s_{t+1}=\mathcal{R}\left(\frac{R(t)\cdot s_Q}{R_{max}} \right),
\end{align}
where $\mathcal{R}(\cdot)$ is the rounding operation, $s_Q$ is the number of quantization level of states, and $R_{max}$ is the maximum available rate from history, i.e., $R_{max}=\max\, R(t-\tau), \ \forall 1\leq\tau\leq M$. After the candidate beam acquirement in Algorithm \ref{Alg2}, we adjust the training power by utilizing the greedy-based mechanism. When a randomized small number $\delta=\left[0, 1\right]$ is higher than a given threshold as $\delta> \delta_{th}$, we select the optimal power action from Q-table as 
\begin{align}\label{Qa}
	a_t = \underset{a'}{\arg\!\max} \ Q(s_t,a'),
\end{align}
whilst a random action is chosen when $\delta\leq \delta_{th}$. Based on the training results in EETBF, we update the Q-table given by
\begin{align} \label{Qup}
	& Q(s_t,a_t)\leftarrow Q(s_t,a_t) + \eta_1\cdot  \notag \\
	& \qquad\qquad \left[ r_t + \eta_2 \cdot \max_{a'}(s_{t+1}, a') - Q(s_t, a_t) \right],
\end{align}
where $\eta_1$ is the learning rate and $\eta_2$ indicates the discount factor. We can infer that $\eqref{Qup}$ leverages both the current beamforming training EE and the potential power policy in the future.

\subsection{Training Beam Initialization}
	We consider that a small amount of offline data for exhaustive beamforming is attainable for channel prediction during the initialization of online learning, which is shown in Algorithm \ref{Alg1}, i.e., $t<M$ indicates insufficient data collection. From the history set of $\boldsymbol{\Xi}(t,M)$, we can estimate the statistical channel state information at the interval $t-\tau$ based on $\eqref{sinr}$ as
\begin{align}
	& H_{i,j}(t-\tau,f,d) = \frac{\gamma_{i,j}(t-\tau)\cdot\sigma^{2}}{P_{i,j}(t-\tau) \cdot G_{h}(\theta_{h,i},\psi_{h,i})\cdot G_{v}(\theta_{v,j},\psi_{v,j})}, \notag \\ 
	& \qquad\qquad\qquad\qquad \forall 1\leq\tau\leq M,
\end{align}
where $\sigma^{2}= N(f,d)+N_{N\!F}$. We employ a power-decaying function to predict the initial channel status rather than using average method due to the compelling significance of more recent data, which is represented by
\begin{align}
	\hat{H}_{i,j}(t,f,d) = \frac{1}{|\mathcal{H}|} \sum_{\tau=1}^{M} \beta^{\tau} \cdot H_{i,j}(t-\tau,f,d),
\end{align}
where $\mathcal{H}=\{H_{i,j}(t-\tau,f,d)\neq 0 | \forall 1\leq \tau \leq M, i \in \boldsymbol{\Phi}_{h}, j\in \boldsymbol{\Phi}_{v} \}$ excludes the unpredictable channels due to potential blocking paths, and $0<\beta< 1$ is the decaying factor. $|\mathcal{H}|$ means the amount of the collected channel information.

After obtaining the estimated beam channels, we now turn to tackle the constraints. We can infer that $\eqref{C5}$ as beam alignment accuracy implies the similar meaning to that in $\eqref{C1}$ and $\eqref{C1-1}$. Therefore, we can rewrite $\eqref{C1}$ and $\eqref{C1-1}$ with alternative constraints respectively as
\begin{align} 
	&\max_{\forall i,j} \ P_{i,j}(t) \cdot\hat{H}_{i,j}(t,f,d) \cdot G_{h}(\theta_{h,i},\psi_{h,i})\cdot G_{v}(\theta_{v,j},\psi_{v,j}) \notag \\
	& \qquad\qquad\qquad\qquad\qquad\qquad 
	\geq \gamma_{th} \cdot \sigma^{2},\label{new1} \\
	& P_{i,j}(t) \cdot\hat{H}_{i,j}(t,f,d) \cdot G_{h}(\theta_{h,i},\psi_{h,i})\cdot G_{v}(\theta_{v,j},\psi_{v,j}) \notag \\
	& \qquad\qquad\qquad\qquad\qquad\qquad
	\geq \gamma_{dec} \cdot \sigma^{2}.\label{new1-1}
\end{align}
Note that the difference between these requirements is that the optimal beam should fulfill $\eqref{new1}$, whereas the other training beams must satisfy the decodability constraint of $\eqref{new1-1}$. Moreover, from $\eqref{C4}$, we know that the maximum number of training beams should not exceed 
\begin{align} \label{new2}
	N_{be} \triangleq |\boldsymbol{\phi}(t)| = \frac{T_{TI}-T_{rq}-T_{fb}-T_{ac}}{T_{bt}}.
\end{align}
Based on $\eqref{new2}$, the top-$N_{be}$ beams with the best possible channel quality in $\hat{\mathcal{H}}=\{\hat{H}_{i,j}(t,f,d)|\forall i \in \boldsymbol{\Phi}_{h}, j\in \boldsymbol{\Phi}_{v}\}$ are selected as the potential training candidates.

%


\begin{table}
\begin{center}
\small
\caption {Parameter Setting of THz}
    \begin{tabular}{ll}
        \hline
        System Parameters & Value \\ \hline 
        BS Serving Radius & $50$ m \\
        System Bandwidth $W$ & $1$ GHz\\
        Operating Frequency $f$ & $[0.1, 1]$ THz \\
        Training power per beam $P_{be}$ & $15$ dBm \\
        Maximum Transmit Power $P_{th}$ & $27$ dBm \\
        Thermal Noise $N_{0}$ & $-174$ dBm/Hz \\
        Sidelobe Beam Gain $\varepsilon$ & $0.1$ \\
        Transmission Interval & $50$ ms\\
        Intervals of $T_{bt},T_{rq}, T_{fb}, T_{ac}$ &  $10$ $\mu$s \\
        Quantization Level $s_{Q}$ & $100$ \\
        Observation Window Size $M$ &  $10$ \\
		Random Action Probability $\delta_{th}$  &  $0.15$ \\
		Learning Rate $\eta_1$ & $0.5$ \\
		Discount Factor $\eta_2$ & $0.5$ \\
        Decaying Factor $\beta$ &  $0.95$ \\
        User Velocity & $\{1.8, 10.8, 27\}$ km/hr\\
        \hline
    \end{tabular} \label{Parameter}
\end{center}
\end{table}

\section{Performance Evaluations} \label{PE}
	
The system performance of proposed EETBF is evaluated through simulations. We evaluate the frequency bands from $0.1$ to $1$ THz, with the corresponding bandwidth of $1$ GHz. The THz channel model in \cite{RRM2} is adopted. Note that the result in \cite{RRM2} implies that the higher frequency bands above $1$ THz suffer from compellingly higher pathloss than those between $\left[0.1,1\right]$ THz. We set the thresholds of training beam SNR to $\gamma_{th}=10$ dB and of the decodable SNR to $\gamma_{th}=0$ dB. The training power per beam is set to $P_{be} = 15$ dBm, while the maximum total power consumption is set to $P_{th}=27$ dBm. The time intervals are set equal to $T_{bt}=T_{rq}=T_{fb}=T_{ac}=10$ $\mu$s. The training duration should be smaller than $T_{TI}=[0.2, 1.2]$ s. Moreover, the training accuracy should be higher than $\xi_{th}=0.7$. The remaining parameter settings are listed in Table \ref{Parameter}. We compare the proposed EETBF with and without training power control as well as the conventional full beam search with full power as baseline. 


\subsection{THz Characteristics}


\begin{figure}[!t]
\centering
	\includegraphics[width=3.2 in]{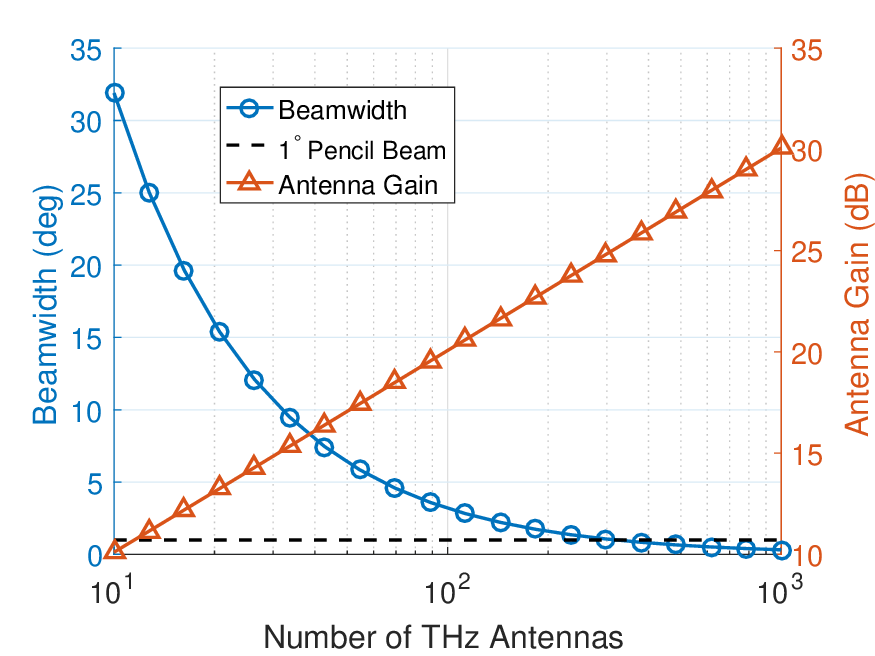}
	\caption{The beamwidth and corresponding antenna gain w.r.t. different numbers of THz antennas.}
	\label{beamg}
\end{figure}	

\begin{figure}[!t]
\centering
	\includegraphics[width=3.5 in]{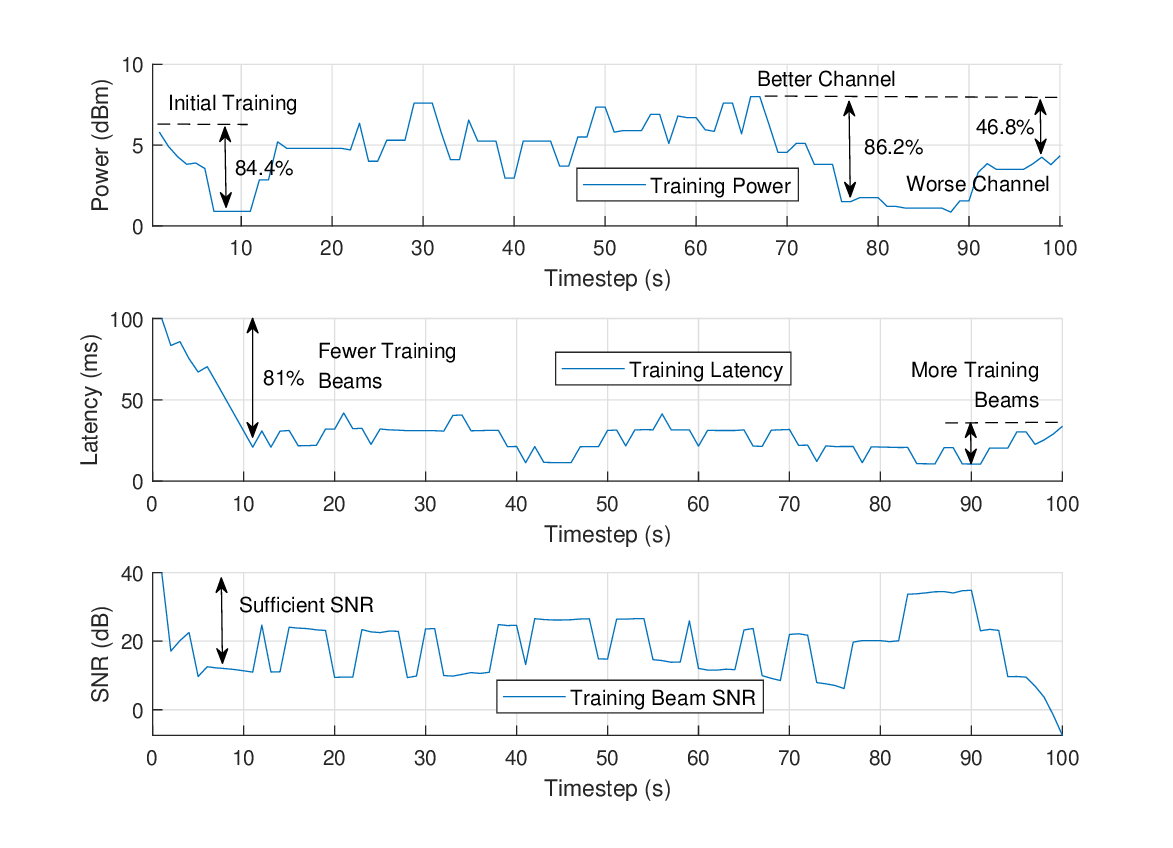}
	\caption{Instantaneous performance of training power, latency and SNR values versus timesteps during the beamforming training process of the proposed EETBF scheme.}
	\label{observe}
\end{figure}	

\begin{figure*}[!t]
\centering
\subfigure[]{\includegraphics[width=3.1 in]{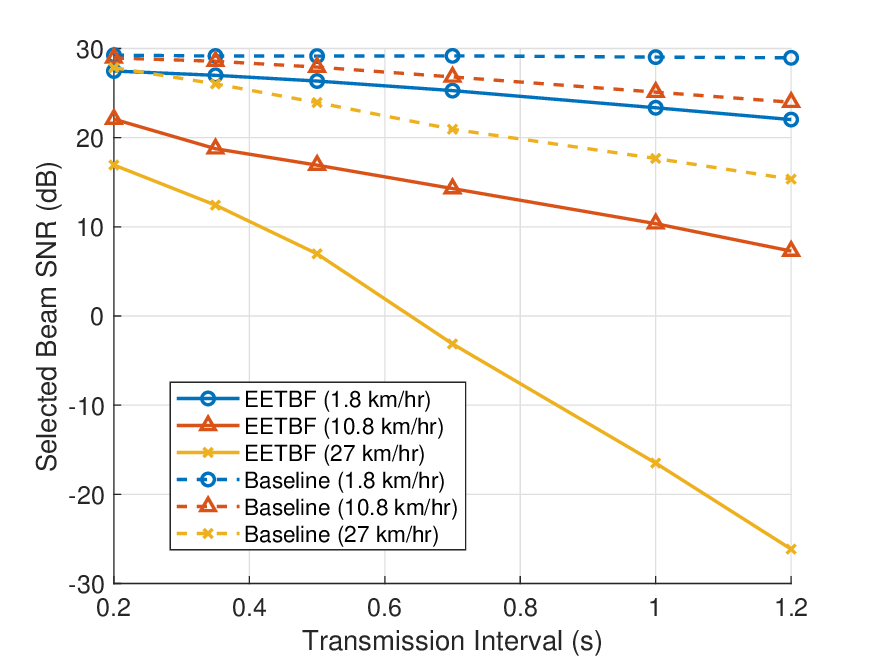} \label{bi1}}
\subfigure[]{\includegraphics[width=3.1 in]{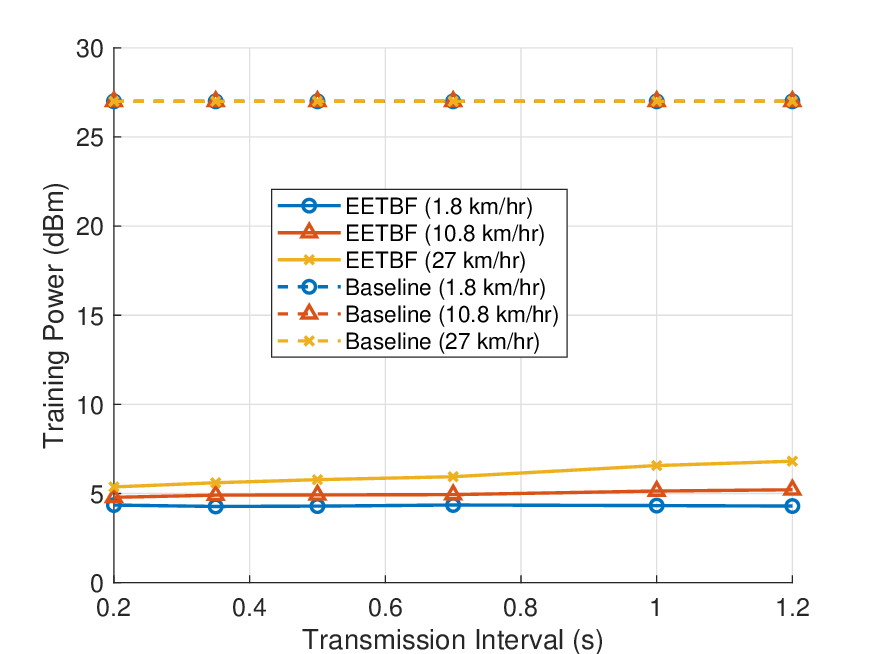} \label{bi2}}
\subfigure[]{\includegraphics[width=3.1 in]{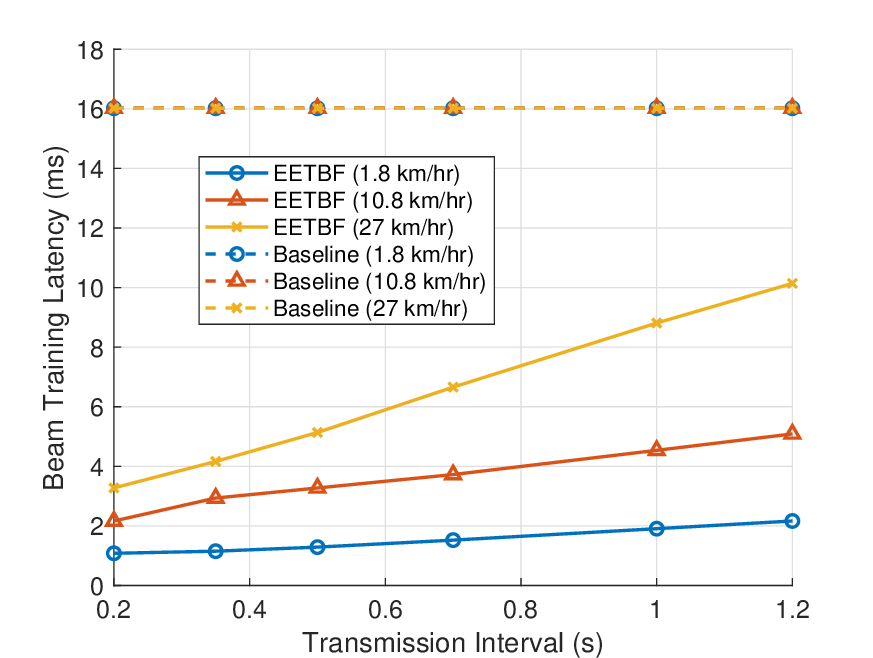} \label{bi3}}
\subfigure[]{\includegraphics[width=3.1 in]{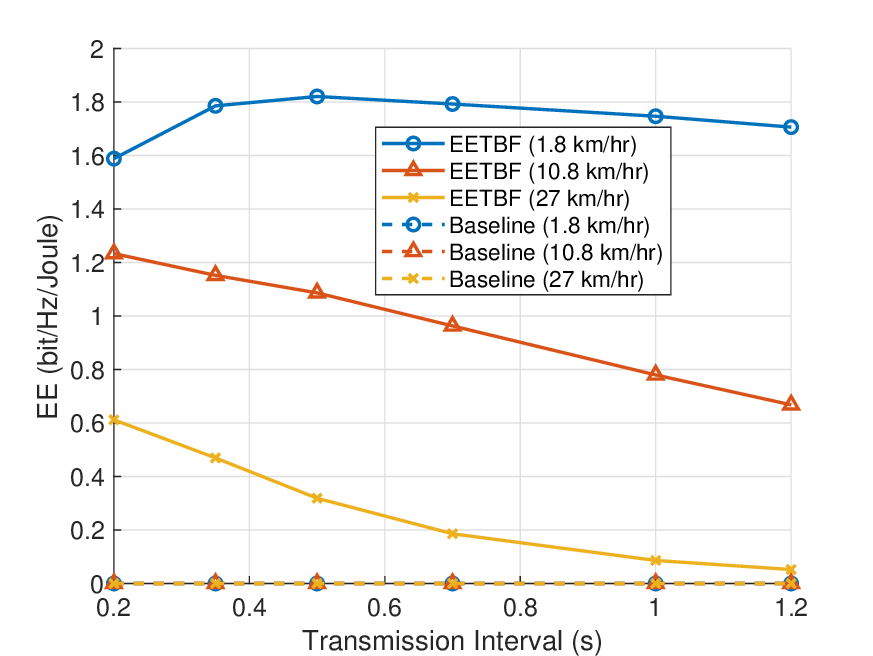} \label{bi4}}
\caption{Performance of EETBF compared to the baseline w.r.t. different velocity values of $\{1.8, 10.8, 27\}$ km/hr and transmission intervals $T_{TI}\in [0.2, 1.2]$ s. (a) Beam SNR (b) Training power (c) Training latency (d) EE.}
\label{bi}
\end{figure*}

In Fig. \ref{beamg}, we study the beamforming requirement of beamwidths and antenna gains with respect to (w.r.t.) different numbers of THz antennas based on \eqref{beamg}. Note that the perfect alignment of mainlobe beam is considered, i.e., the transmit beam angle $\psi_{\mathcal{A},x}$ falls into the boresight angles $\tilde{\psi}_{\mathcal{A}}$ of the THz BS. We can infer that the beamwidth is inversely proportional to the number of antennas, whereas antenna gain is linearly proportional to  the number of antennas. That is, a smaller beamwidth can generate a higher beam gain, but more antennas are required for the beamforming. As shown in Fig. \ref{beamg}, $30$ dB beamforming gain requires around a thousand antennas. Moreover, around $500$ antennas are needed to form a $1^{\circ}$ pencil-beam. 

\begin{figure*}[!t]
\centering
\subfigure[]{\includegraphics[width=3.2 in]{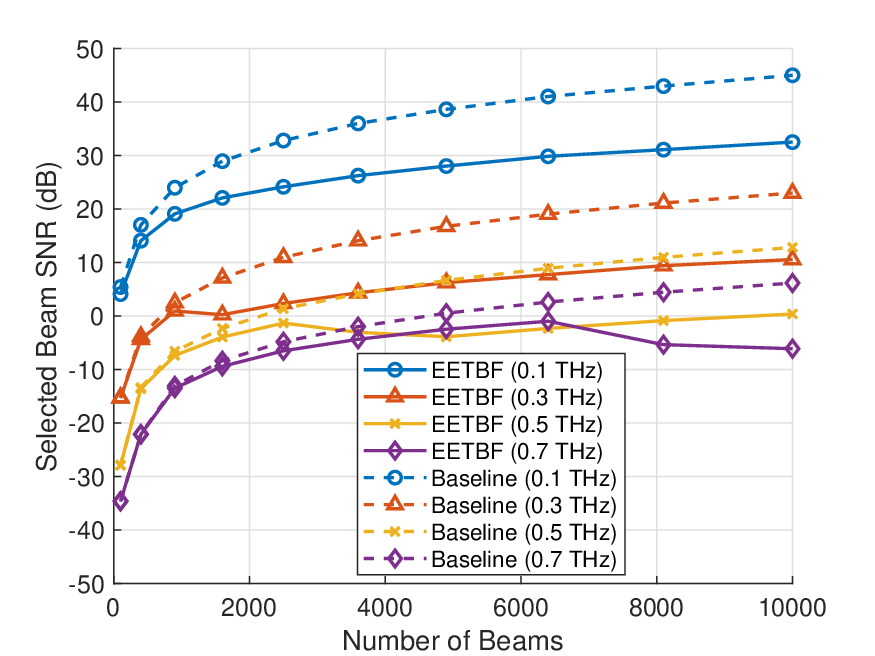}\label{fr1}}
\subfigure[]{\includegraphics[width=3.2 in]{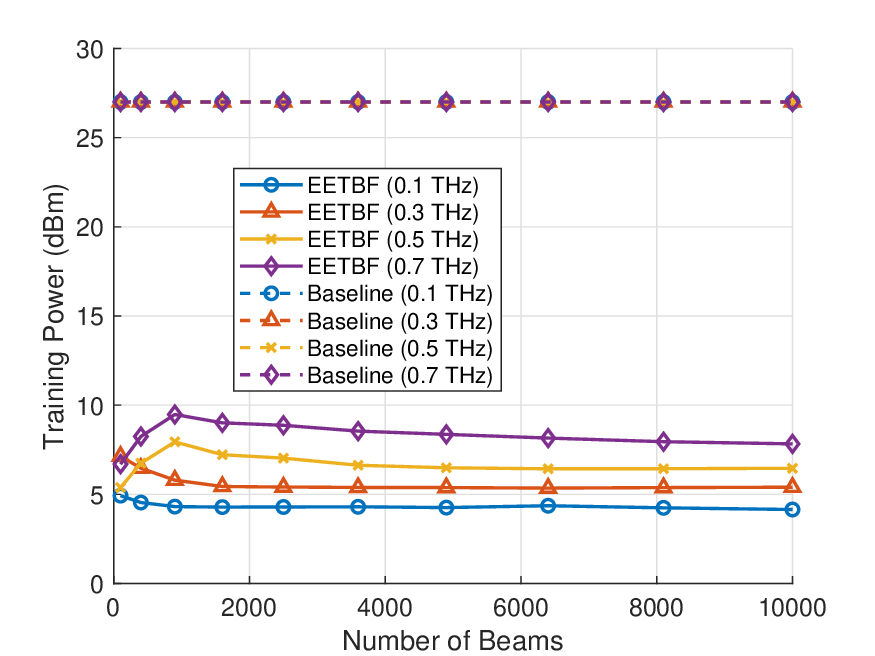}\label{fr2}}
\subfigure[]{\includegraphics[width=3.2 in]{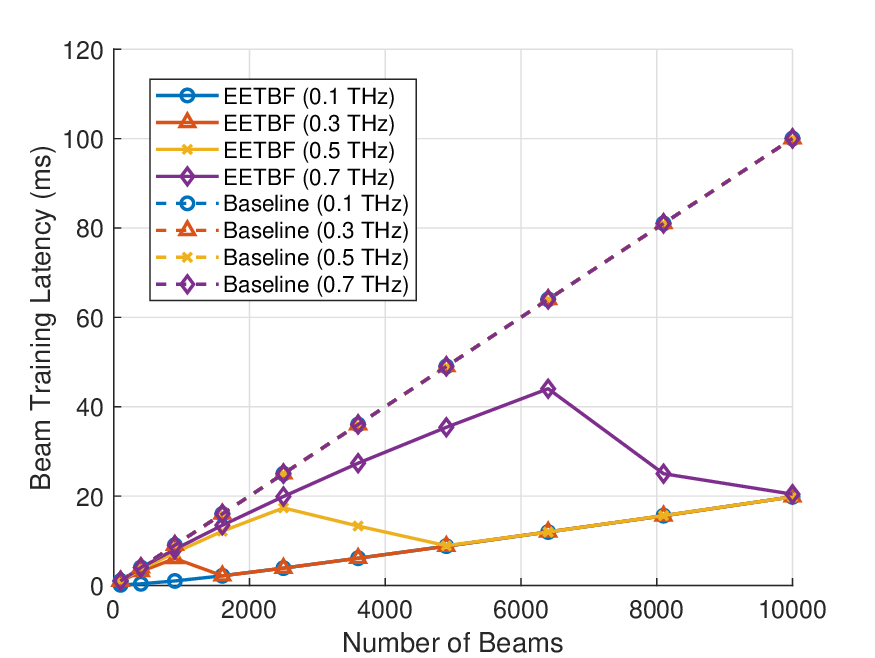}\label{fr3}}
\subfigure[]{\includegraphics[width=3.2 in]{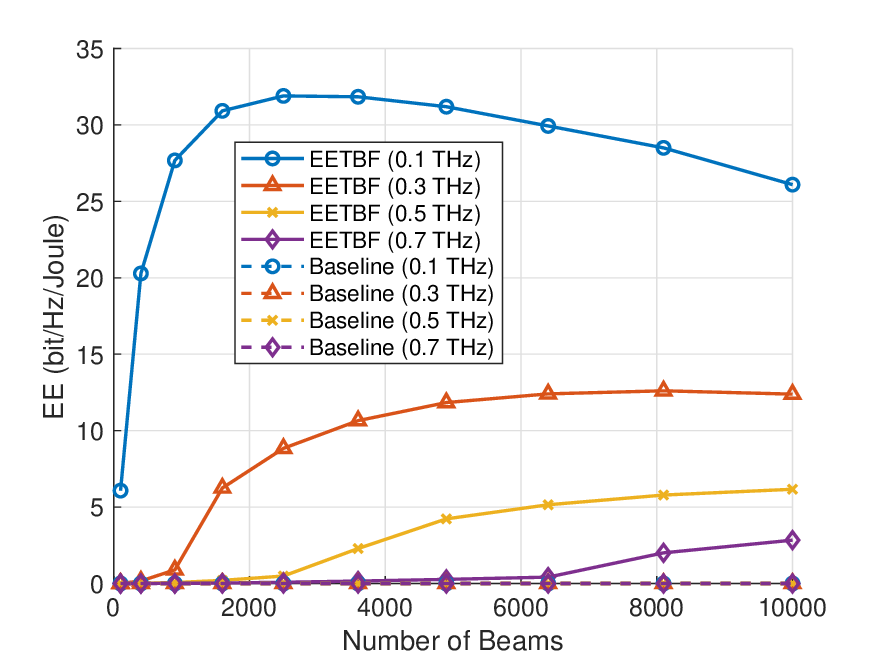}\label{fr4}}
\caption{Performance of EETBF compared to the baseline w.r.t. different frequencies of $\{0.1, 0.3, 0.5, 0.7\}$ THz and numbers of beams $[100, 10000]$. (a) Beam SNR (b) Training power (c) Training latency (d) EE.}
\label{fr}
\end{figure*}

In Fig. \ref{observe}, we analyze the instantaneous performances for different stages of beamforming training of the proposed EETBF scheme. We set the transmission interval to $T_{TI}=1$ s for better observation. The training power, training latency, and the corresponding SNR values are studied. We can observe that the highest number of beams is required at the beginning owing to the initial training stage using exhaustive search. However, at a timestep of $10$ s, EETBF can preserve $84.4\%$ training power and reduce training latency by $81\%$. This is due to the fact that it requires fewer but more energy-efficient beams after learning the potential information of user trajectory. Moreover, an SNR of around $10$ dB is attained for sufficiently decodable signals. When low-latency training is sustained under user mobility from $10$ s to $65$ s, EETBF potentially requires more power to compensate for weaker beams due to uncertain user movements. With better channel quality from the timestep of around $70$ s to $90$ s, around $86.2\%$ of power can be preserved for high energy efficiency. However, from $90$ to $100$ s compared to that at $65$ s, only around $46.8\%$ of power is preserved under worse channel conditions. Such case also provokes more training beams involved for finding those with better signal quality.

\begin{figure*}[!t]
\centering
\subfigure[]{\includegraphics[width=3.2 in]{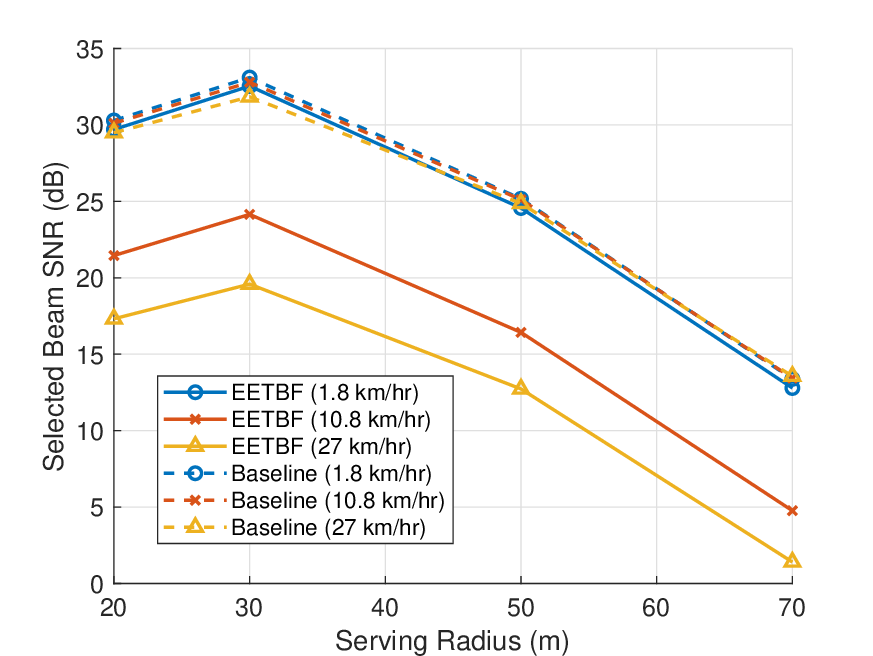}\label{ran1}}
\subfigure[]{\includegraphics[width=3.2 in]{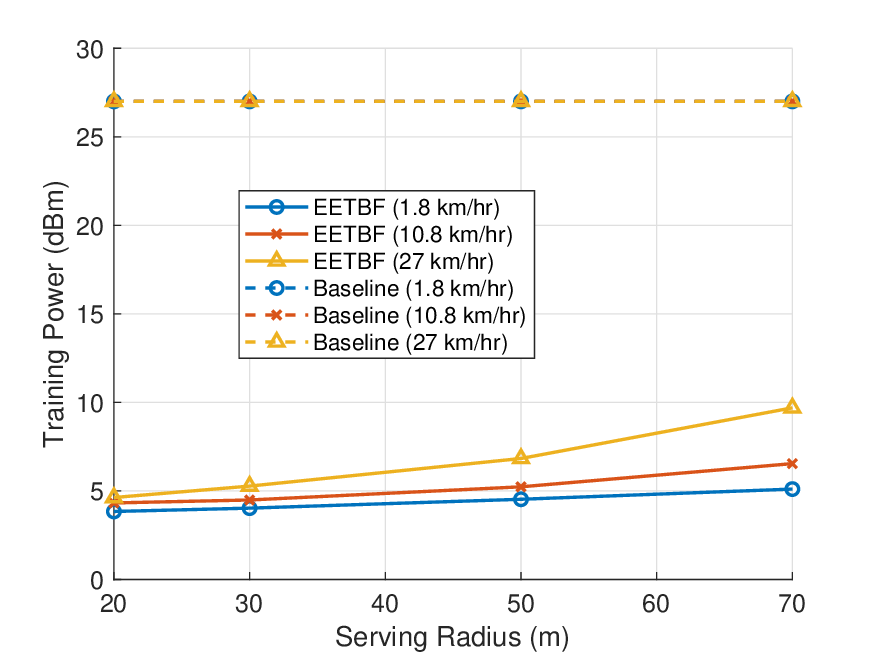}\label{ran2}}
\subfigure[]{\includegraphics[width=3.2 in]{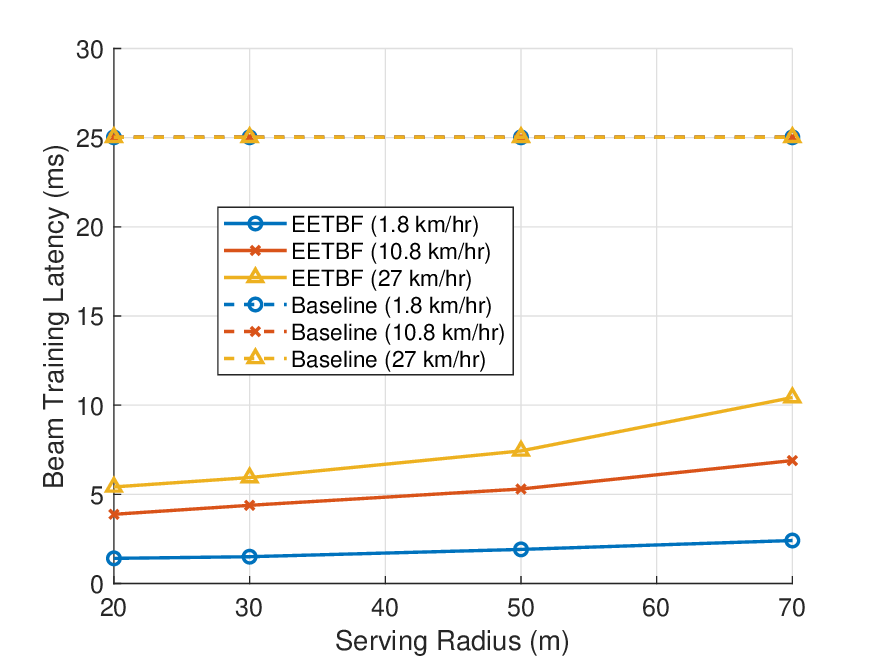}\label{ran3}}
\subfigure[]{\includegraphics[width=3.2 in]{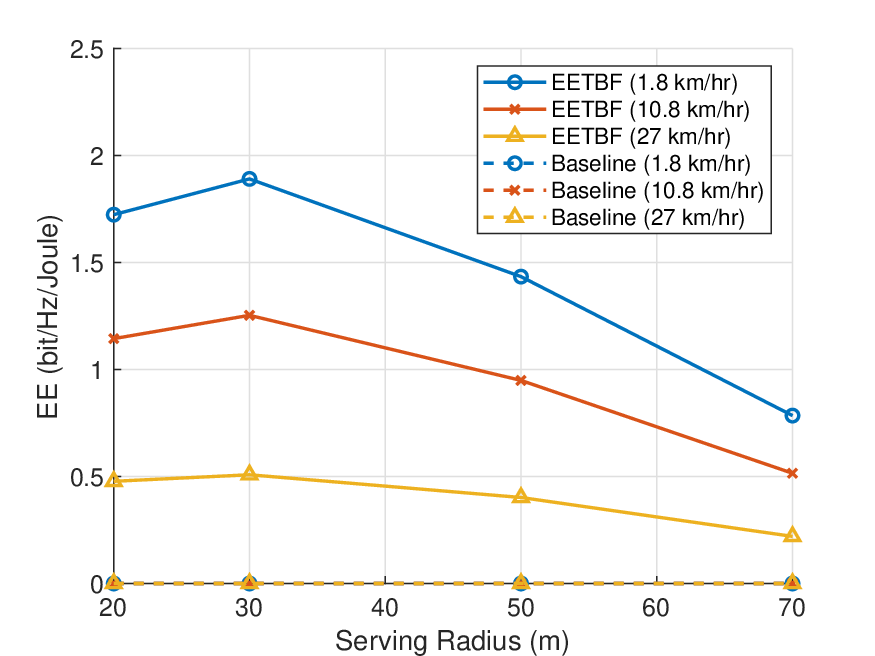}\label{ran4}}
\caption{Performance of EETBF compared to the baseline w.r.t. velocity values of $\{1.8,10.8,27\}$ km/hr and serving radius of $[20, 70]$ m. (a) Beam SNR (b) Training power (c) Training latency (d) EE.}
\label{ran}
\end{figure*}

\subsection{Effect of Transmission Intervals}

In Fig. \ref{bi}, we investigate the performance of EETBF in comparison with the baseline schemes w.r.t. different velocity values of $\{1.8, 10.8, 27\}$ km/hr and transmission intervals of $T_{TI}\in [0.2, 1.2]$ s. It is worth mentioning that $1.8$ km/hr simulates random pedestrian walk, whereas $10.8$ and $27$ km/hr are for vehicular speeds. Note that the baseline scheme considered is the conventional full beam search with full power during beam training. In Fig. \ref{bi1}, we can observe that a higher velocity induces a lower beam SNR, since faster movement potentially leads to the expired training beams. Furthermore, with longer transmission intervals, out-of-date beams also cause low SNR values, whereas shorter transmission intervals can frequently refresh up-to-date beams for acquiring high SNR. Thanks to the full beam search, the highest beam SNR can be obtained. However, as illustrated in Fig. \ref{bi2}, it requires the highest power consumption of $27$ dBm compared to the proposed EETBF of around $5$ dBm. A slightly higher power of EETBF is required with higher transmission interval in order to obtain higher SNR to compromise high latency, as depicted in Fig. \ref{bi3}. Additionally, as shown in Fig. \ref{bi3}, the baseline scheme has the highest beam training latency of $16$ ms. We can also observe that the proposed EETBF results in higher latency with the increment of transmission intervals, which is the result that more beams are required to retrain beamforming due to the out-of-date historical beam data. Combining Figs. \ref{bi1} to \ref{bi3} provides the total result of EE in Fig. \ref{bi4}. We can infer that it reaches the optimal transmission interval of $0.5$ s for pedestrian velocity. This is because frequent beam training consumes more energy compared to longer transmission interval, whilst high latency and low SNR metrics dominate longer transmission interval. However, more beams and power are required for training, which results in a decreasing EE. Although the optimal beam can  be obtained by the full beam search, its comparatively higher power consumption and beam training latency provoke the near zero EE.

\begin{figure*}[!t]
\centering
\subfigure[]{\includegraphics[width=3.2 in]{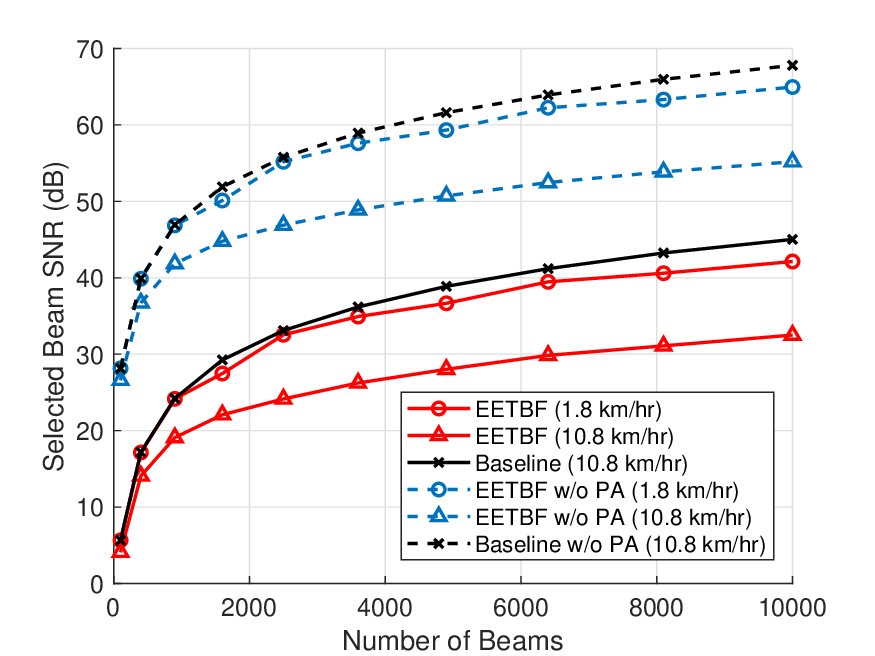}\label{bm1}}
\subfigure[]{\includegraphics[width=3.2 in]{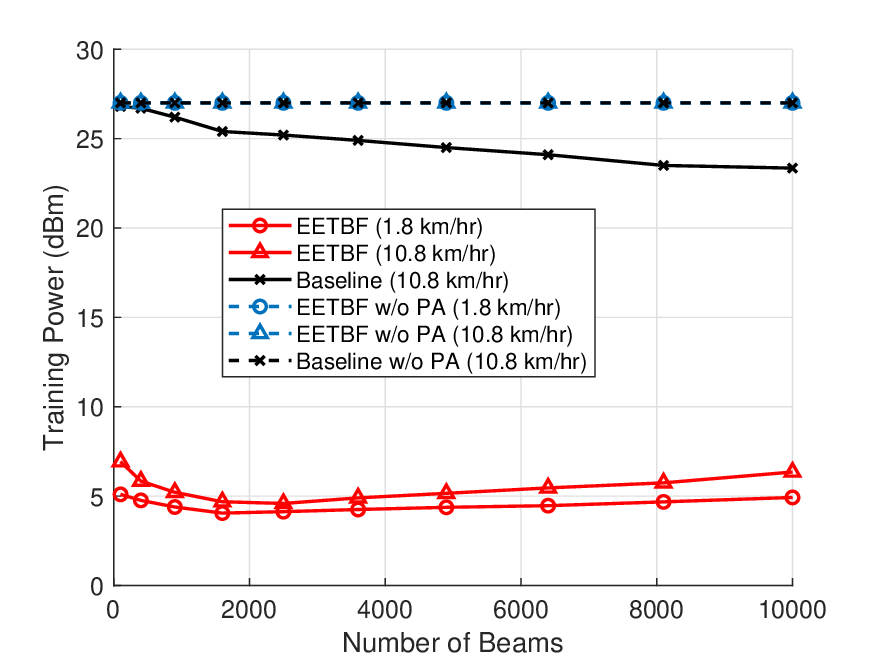}\label{bm2}}
\subfigure[]{\includegraphics[width=3.2 in]{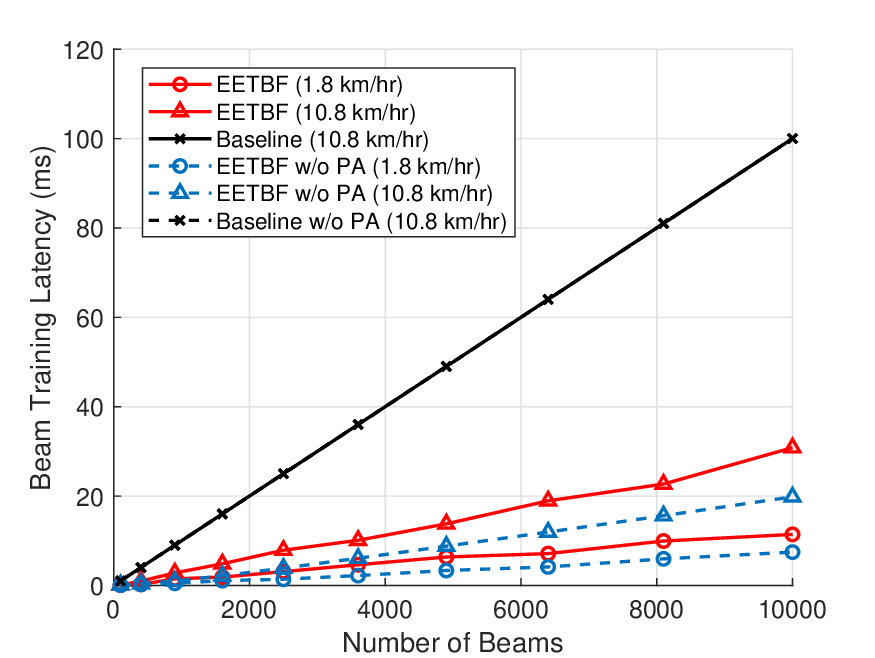}\label{bm3}}
\subfigure[]{\includegraphics[width=3.2 in]{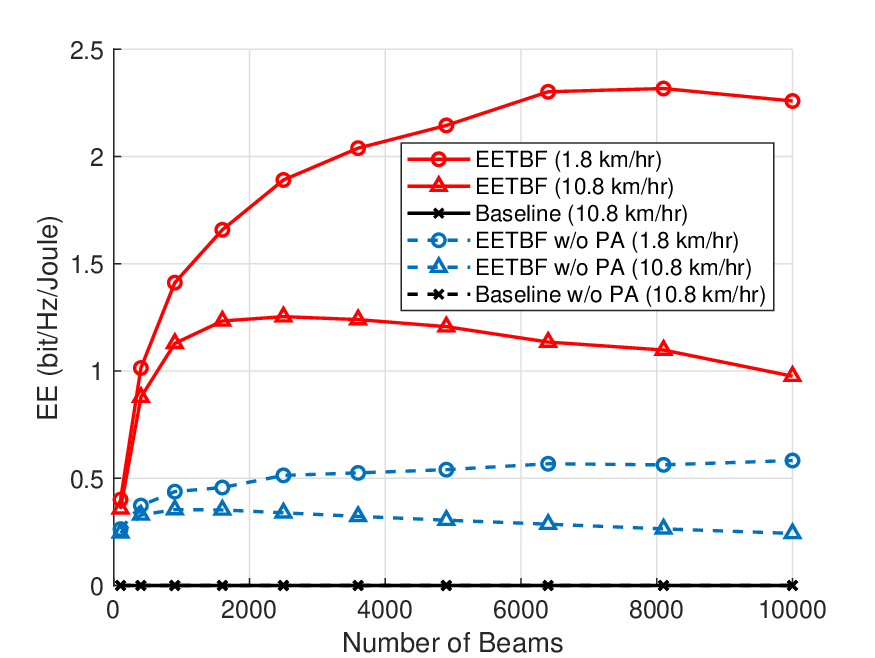}\label{bm4}}
\caption{Comparison of EETBF compared to that without PA mechanism and the baseline under different velocity values of $\{1.8, 10.8\}$ and numbers of beams $[100, 10000]$. (a) Beam SNR (b) Training power (c) Training latency (d) EE.}
\label{bm}
\end{figure*}

\subsection{Impact of Operating THz Bands}

In Fig. \ref{fr}, we study the performance of EETBF and compare it with the baseline scheme, when different frequencies of $\{0.1, 0.3, 0.5, 0.7\}$ THz and numbers of beams $[100, 10000]$ are considered. Note that a larger number of beams correspond to a smaller beamwidth, which can refer to Fig. \ref{beamg}. In Fig. \ref{fr1}, we can know that in most cases, more beams can achieve a higher SNR due to its higher degree of freedom of beam training. Nevertheless, with the increase of the number of beams at high frequency bands, lower SNR obtained may be affected by reduced training power and high latency, due to the fact that more beams are required to be trained for achieving high EE. 
As a result, we can see different trends of EETBF from Fig. \ref{fr2}. 
In Fig. \ref{fr2}, with a higher operating frequency of $0.7$ THz, it consumes more power to train fewer beams to compensate high pathloss, but with less power for enormous beams to compromise high training latency. By contrast, the frequencies of $\{0.1, 0.3\}$ THz with lower pathloss can use more power to obtain high-quality beams with low latency. Observing from Fig. \ref{fr3}, we have intriguingly turning points when $\{900, 2500, 6400\}$ beams at frequencies of $\{0.3, 0.5, 0.7\}$ THz, respectively. This is because it tends to sacrifice some training beams in exchange of more data transmission time even a low-SNR beam is obtained. Another turning points can be found when there are $\{1600, 4900\}$ beams at frequencies of $\{0.3, 0.5\}$ THz, respectively. With lower training power consumed, more beams are still required to obtain a moderate SNR. In Fig. \ref{fr4}, we can know that the baseline scheme attains nearly zero EE due to the high power and the linearly increasing latency. The compellingly low beam SNR dominates the result at higher THz frequencies even with low latency, leading to near zero EE. This implies that more beams are needed to generate high-SNR beams to compensate for the substantially high THz pathloss. However, under the modest pathloss, the EE performance curve of proposed EETBF has a concave characteristic with an optimal number of beams at $0.1$ THz, owing to the fact that our EETBF can strike a compelling tradeoff between beam SNR, training power and training latency.

\subsection{Serving Radius}

In Fig. \ref{ran}, we evaluate the performance of EETBF and compare it with the performance of the baseline scheme, when different frequencies of $\{0.1, 0.3, 0.5, 0.7\}$ THz and serving radius of $[20, 70]$ m are considered. In Fig. \ref{ran1}, we can observe that there exists an optimal serving radius of about $30$ m for all cases. When the serving radius is below $30$ m, more power and more training beams are leveraged to obtain a higher beam SNR. On the contrary, when the serving radius is more than $30$ m, the substantial pathloss dominates the result, leading to a lower beam SNR even when much more power and many beams for training are used. Accordingly, the EE has the asymptotic curves as that of beam SNR, and the optimal EE can be obtained when the serving radius is about $30$ m.

\subsection{Effect of Power Control}

In Fig. \ref{bm}, we compare the proposed EETBF scheme with and without PA mechanism as well as the baseline scheme at the velocity of $\{1.8, 10.8\}$ km/hr and the different numbers of beams of $[100, 10000]$. We can observe from Fig. \ref{bm1} that the schemes with power control attain a lower beam SNR in exchange for preserving more power as depicted Fig. \ref{bm2}. However, it consumes more power at higher velocity to compromise the low SNR. Owing to the high latency induced from full beam search, the training power of the baseline scheme with power control is declined in order to preserve power resources. Moreover, the training power of EETBF presents the convex curve shape. Less power is utilized for attaining higher EE with increasing numbers of beams, whereas more power is required to train more beams. Therefore, as illustrated in Fig. \ref{bm3}, the EEFBT with power control has a slightly higher latency compared to that with full power utilization. In Fig. \ref{bm4}, we can see that the EETBF with power control achieves the highest EE, which is about three times higher than that of the EETBF without power control. By contrast, even the baseline scheme makes use of power conservation, it still provokes zero rate as well as zero EE due to the compellingly high beam training latency.

\begin{table*}[!t]
\caption{Benchmark Comparison}
\renewcommand{\arraystretch}{1.2}
\scriptsize
\resizebox{\textwidth}{!}{
\begin{tabular}{|l|c|c|c|c|c|c|c|c|}
\hline
& \begin{tabular}[c]{@{}c@{}}Protocol or\\ Framework\end{tabular} & \begin{tabular}[c]{@{}c@{}}Feedback\\ Required\end{tabular} & \begin{tabular}[c]{@{}c@{}}THz/mmWave \\ Compatibility \end{tabular} & \begin{tabular}[c]{@{}c@{}}Latency \\ (ms)\end{tabular} & \begin{tabular}[c]{@{}c@{}}SNR \\ (dB)\end{tabular} & \begin{tabular}[c]{@{}c@{}}Rate \\ (bps/Hz)\end{tabular} & \begin{tabular}[c]{@{}c@{}}Power \\ (dBm)\end{tabular} & \begin{tabular}[c]{@{}c@{}}EE \\ (bit/Hz/Joule)\end{tabular} \\ \hline
\textbf{Proposed EETBF}
& \multirow{3}{*}{This   work}                                                                               & \multirow{7}{*}{\textbf{Once}} 
& THz     & 5                                                       & 25                                                  
& 7.5                                                      & \textless 8                                            & 2                                                            \\ \cline{1-1} \cline{4-9} 
\begin{tabular}[c]{@{}l@{}}\textbf{Proposed EETBF} \\ w/o PA Mechanism\end{tabular}        &                                                                                                            &                                                             & THz     & 5                                                       & 40                                                  & 12                                                       & 27   (Max)                                             & 0.5                                                          \\ \cline{1-2} \cline{4-9} 
Exh. w/ PA                                                                         & \multirow{2}{*}{\begin{tabular}[c]{@{}c@{}}   802.15.3d  \cite{IEEE3d}  \\ 802.11ad/ay \cite{IEEEaday} \end{tabular}} &                                                             & Both    & \textgreater 50                                         & 30                                                  & $\sim$0                                                  & \textless   8                                          & $\sim$0                                                      \\ \cline{1-1} \cline{4-9} 
Exh. w/o PA &                                                                                                            &                                                             & Both    & \textgreater 50                                         & 50                                                  & $\sim$0                                                  & 27   (Max)                                             & $\sim$0                                                      \\ \cline{1-2} \cline{4-9} 
\begin{tabular}[c]{@{}l@{}}Random   Search\\ ($\sim$1000   beams)\end{tabular} & Arbitrary                                                                                                       &                                                             & Both    & 10                                                      & 10                                                  & 3                                                        & 27   (Max)                                             & 0.1                                                          \\ \hline
Iterative Search                                                               & \multirow{3}{*}{\begin{tabular}[c]{@{}c@{}} 802.15.3d \cite{RIS1}\\       802.11ad \cite{mmWTree} \end{tabular}}       & \multirow{5}{*}{Per   beam}                                 & Both    & 15                                                      & 25                                                  & 6                                                        & 27   (Max)                                             & 0.2                                                          \\ \cline{1-1} \cline{4-9} 
\begin{tabular}[c]{@{}l@{}}Iterative Search  \\ (Power Control)\end{tabular}  &                                                                                                            &                                                             & THz     & 10                                                      & 20                                                  & 5                                                        & \textgreater 27                                        & 0.15                                                         \\ \cline{1-2} \cline{4-9} 
Linear Search                                                                  & \multirow{2}{*}{   802.15.3c \cite{mmWLBS}}                                                                        &                                                             & mmWave     & 35                                                      & 20                                                  & 2                                                        & 27   (Max)                                             & 0.07                                                         \\ \cline{1-1} \cline{4-9} 
Binary search                                                                  &                                                                                                            &                                                             & mmWave     & 20                                                      & 25                                                  & 4                                                        & 27   (Max)                                             & 0.15                                                         \\ \hline
\end{tabular}\label{tablecom}}
\end{table*}

\subsection{Benchmark Comparison}

In Table \ref{tablecom}, we compare the proposed EETBF scheme with some existing methods in open literature, including the conventional exhaustive beam training \cite{IEEE3d,IEEEaday}, random beam search, iterative search \cite{RIS1,mmWTree} and the linear/binary search methods \cite{mmWLBS}. In random search, we randomly select around $1000$ beams for training. The iterative search employs a tree-based approach, starting with the largest candidate beamwidth and progressively reducing the beamwidth. Furthermore, both linear and binary search methods utilize a fixed beamwidth. Linear search trains the consecutive neighboring beams, whereas binary search selects the intersecting beam sectors. The corresponding adopted protocols of benchmarks for implementation are also listed in Table \ref{tablecom}. We can observe that the proposed EETBF as well as the exhaustive/random search require only a single feedback, which is carried out after all the training beams from BS are received. These are applicable in IEEE 802.11ad/ay frameworks and protocols. However, the mechanisms in \cite{RIS1,mmWTree,mmWLBS} require frequent beam reports, which potentially lead to much higher training latency compared to the proposed scheme. Such methods are applicable for per-beam feedback based IEEE 802.15.3c or 802.15.3d frameworks and protocols. Theoretically, per-beam feedback has around twice the latency of single feedback. For instance, the EETBF has a latency of about $5$ ms, whilst the linear and binary search have the respective latency of about $35$ ms and $20$ ms. Moreover, with the compellingly high latency induced by full beam search, no data transmission interval is left, leading to nearly zero rate as well as EE performance. Although high-SNR beams can be obtained through full power utilization, the EETBF without PA has an EE of only $0.5$ bit/Hz/Joule. We further notice that the iterative search requires power higher than the maximum allowable power threshold, as it commences its training with wider beamwidths. In other words, much more power should be used to compensate for the lower beam gain at the beginning of the training; otherwise, no beams will be obtained. However, the proposed EETBF utilizes equal small beamwidths for beam training, which alleviates the issue. Moreover, the high latency of both linear and binary search methods is attained owing to the miscellaneous beam training and excessive feedbacks. To conclude, the proposed EETBF with PA can compromise the tradeoff between effective rate and power consumption, resulting in the highest EE of 2 bit/Hz/Joule.

\section{Conclusions and Future Works} \label{CON}

\subsection{Conclusion Remarks}	
	In this work, we have conceived a joint EE-oriented THz 3D beamforming training, which includes the EETBF-BT for beam training and the EETBF-PA for power control. Owing to the high pathloss at THz frequencies, our EETBF is capable of striking a compelling tradeoff between the beamforming gain and power consumption for achieving the highest EE performance. Simulation results have revealed the optimal configurations of the transmission interval, number of beams as well as the serving radius under various user's moving velocities and frequency bands. The studies and results have shown that the proposed EETBF outperforms the existing benchmarks of leveraging full beam search, iterative search, linear/binary search as well as the non-power-control based mechanism found in open literature. The proposed EETBF can achieve the lowest training latency and power consumption, resulting in the highest effective rate and EE performance.

\subsection{Challenges and Future Works}	

Here, we provide the potential challenges and real-world feasibility of implementing EETBF in existing THz communication systems. First, a more practical beamforming model can be considered, which may affect the 3D beamforming gain and its alignment. The uncertain misalignment will affect the training policy, especially under high mobility scenarios, potentially leading to lost feedback information. Second, out-of-band frequency utilization can be adopted to significantly enhance the training efficiency. For example, beam training using hybrid bands in sub-6 GHz or mmWave bands can transfer their partial knowledge to THz bands. However, such a mechanism might complicate the framework and protocol, which can be regarded as an open research. Third, existing architectures allow the BS to control its power in different sub-channels or sub-carriers mostly only for data transfer. Therefore, BS requires more complex power control for each training beam and data beam in EETBF. To elaborate a little further, we can also regard the proposed framework and EETBF as a general mechanism for compromising single- and per-beam based feedback. As an extension, training performance can be improved by appropriately adjusting the total frame interval and the number of training beams as well as of feedbacks based on the dynamic environments. Such framework requires a more complex algorithm and protocol, which can be left as the future promising work.

\bibliographystyle{IEEEtran}
\bibliography{IEEEabrv}

\end{document}